\documentclass[preprint,aps,12pt,showpacs,nofootinbib,tightenlines]{revtex4}
\usepackage{amsmath}
\usepackage{graphicx}
\usepackage{amssymb}

\begin{document}

\title{Consistency of Loop Regularization Method and \\
Divergence Structure of QFTs Beyond One-Loop Order }
\author{Da Huang}
\email{dahuang@phys.nthu.edu.tw}
\affiliation{Kavli Institute for Theoretical Physics China (KITPC) at the CAS \\
State Key Laboratory of Theoretical Physics (SKLTP) \\
Institute of Theoretical Physics, Chinese Academy of Science, Beijing,100190
\\
Department of Physics, National Tsinghua University, Hsinchu, Taiwan 300}
\author{Ling-Fong Li}
\email{lfli@cmu.edu}
\affiliation{Kavli Institute for Theoretical Physics China (KITPC) at the CAS \\
Department of Physics, Carnegie Mellon University, Pittsburgh, PA 15213}
\author{Yue-Liang Wu}
\email{ ylwu@itp.ac.cn}
\affiliation{Kavli Institute for Theoretical Physics China (KITPC) at the CAS \\
State Key Laboratory of Theoretical Physics (SKLTP) \\
Institute of Theoretical Physics, Chinese Academy of Science, Beijing,100190
\\
University of Chinese Academy of Sciences}
\date{\today }

\begin{abstract}
We study the problem how to deal with tensor-type two-loop integrals in the
Loop Regularization (LORE) scheme. We use the two-loop photon vacuum
polarization in the massless Quantum Electrodynamics (QED) as the example to
present the general procedure. In the processes, we find a new divergence
structure: the regulated result for each two-loop diagram contains a
gauge-violating quadratic harmful divergent term even combined with their
corresponding counterterm insertion diagrams. Only when we sum up over all
the relevant diagrams do these quadratic harmful divergences cancel,
recovering the gauge invariance and locality.
\end{abstract}

\maketitle

\hfill


~\newline
~\newline
~\newline

\section{Introduction}

This is the second of a series of papers to study the consistency of Loop
Regularization (LORE) scheme and the general structure of QFT models in the
higher-loop calculations. In our first paper\cite{Huang}, we have shown that
the LORE method is consistent for general scalar-type two-loop integrals. We
studied the general $\alpha \beta \gamma $ integrals\cite{'tHooft:1972fi}
and mainly solved the problem of disentangling the overlapping divergences,
which is the main new feature of the perturbative calculations beyond
one-loop order. We argued that the overlapping divergence structure would,
in general, transforms some of the ultraviolet (UV) divergences into the
integrals of the ultraviolet divergence-preserving(UVDP) parameters (or
Feynman parameters). We found that the LORE method was particularly suitable
to regularize the UV divergences of this kind since we can easily transform
the UVDP parameter integrals into the irreducible-loop-integral(ILI)-like
integrals by multiplying the integration variable by a mass scale, which is
exactly the object regulated in the LORE method. Furthermore, it was very
useful to introduce the Bjorken and Drell's analogy between the Feynman
diagrams and electric circuits, which allowed us to identify the UV
divergences of the UVDP parameters with those contained in the subdiagrams
of the original Feynman diagrams. Then we applied all the tools we found to
the study of two-loop calculations of the $\phi ^{4}$ model and obtained the
consistent results for $\beta $-function of the coupling constant and mass
parameter. Finally, we gave the general procedure of the application of the
LORE method to the higher-loop calculations, even beyond two-loop order.

In the present paper, we will focus on one of the important remaining
problems-- how to deal with the tensor-type integrals at higher-loop order.
Recall that at one-loop order in LORE tensor-type ILIs and scalar-type ILIs
can be related by introducing consistency conditions, which relate these
type of integrals with the same divergent behavior. So the immediate
question is whether these consistent conditions, which should be applied to
the 1-folded ILIs of each loop momentum variable, is still consistent with
the general features we find in our previous paper. In particular, we want
to ask whether the consistent conditions can guarantee the cancellation of
various harmful divergences. Another question involves the gauge invariance
which is ensured by the consistency conditions in LORE as shown in \cite%
{wu1,wu2}. Thus, it is necessary to inquire whether the consistency
conditions can still preserve gauge invariance and the associated Ward
identities. In order to study these questions, we use the massless Quantum
Electrodynamics (QED) as our simplest example, which involves the
tensor-type integrals and the gauge invariance at the same time.
Specifically, we shall calculate the two-loop photon vacuum polarization
diagrams in detail. As will be shown in the following calculation, the LORE
method, with the prescription of consistency conditions, can give the
sensible results which satisfy all the requirements: gauge invariance and
locality. We should emphasize that the result above only requires the use of
consistency conditions found at one-loop order and does not need to
introduce additional two-loop order consistency conditions.

One special feature worth mentioning is the appearance of a new divergence
structure: each two-loop Feynman diagram contains a quadratically harmful
divergence even when we combine the result with the corresponding one-loop
counterterm insertion diagrams. These new divergence structure is expected
and compatible with the usual power counting for each diagrams. However,
these quadratically harmful divergences will cancel with each other when we
sum up all three group of diagrams in this order, rending the final result
local. Another way to see this cancellation can be obtained from the
requirement of gauge invariance of the final result since these
quadratically harmful divergences break the gauge invariance and must
disappear in the final result. We argue that this is a general feature for
gauge theories, since quadratic divergences would in general give the photon
a mass, which is incompatible with gauge invariance. The appearance of this
new divergence structure explicitly shows one of the main advantage of LORE
method that the LORE method enables us to calculate the full result for any
Feynman diagram, especially the quadratic divergences, which is absent in
many other regularization methods, like the popular dimensional
regularization (DR). In the light of detailed and complete calculations, we
conclude that the LORE method can properly regularize tensor-type high-loop
integrals (at least up to two-loop) and consistently apply to general gauge
theories, like the Standard Model (SM) of particle physics.

The paper is organized as follows. Section 2 gives a short introduction to
the LORE method. In Section 3, we first compute all of the one-loop
one-particle-irreducible diagrams that need to be regulated, in order to set
our notations. Then we present the results of the massless QED vacuum
polarization at two-loop order by using the LORE method, leaving the many of
the details to the Appendix B and C. Appendix A contains some useful
formulae in the LORE method. Appendix D contributes to the detailed
calculation of a new doubly-logarithmic divergent integral encountered in
our two-loop calculations. 

\section{Irreducible Loop Integrals(ILIs) and the Prescription of Loop
Regularization (LORE)}

In this section, we give a brief introduction to the Loop Regularization
(LORE). A more detailed discussion will be found in the seminal paper by one
of the authors \cite{wu1}.

It has been shown in\cite{wu1, wu2} that all one loop Feynman integrals can
be reduced into the following 1-fold ILIs by using the standard Feynman
parameterization method:
\begin{eqnarray}
I_{-2\alpha } &=&\int \frac{d^{4}k}{(2\pi )^{4}}\frac{1}{(k^{2}-M^{2})^{2+%
\alpha }},  \notag \\
I_{-2\alpha \ \mu \nu } &=&\int \frac{d^{4}k}{(2\pi )^{4}}\frac{k_{\mu
}k_{\nu }}{(k^{2}-M^{2})^{3+\alpha }},\hspace{8mm}\alpha =-1,0,1,2,...
\notag \\
I_{-2\alpha \ \mu \nu \rho \sigma } &=&\int \frac{d^{4}k}{(2\pi )^{4}}\frac{%
k_{\mu }k_{\nu }k_{\rho }k_{\sigma }}{(k^{2}-M^{2})^{4+\alpha }}
\end{eqnarray}%
with $I_{2}$ and $I_{0}$ corresponding to the quadratic and logarithmic
divergent integrals. Here the effective mass factor $M^{2}$ is a function of
the external momenta $p_{i}$, the masses of particles $m_{i}$ and the
Feynman parameters.

When the regularized 1-fold ILIs satisfy the following consistency conditions%
\cite{wu1, wu2}:
\begin{eqnarray}  \label{cc}
& & I_{2\mu\nu}^R = \frac{1}{2} g_{\mu\nu}\ I_2^R, \quad
I_{2\mu\nu\rho\sigma }^R = \frac{1}{8} (g_{\mu\nu}g_{\rho\sigma} +
g_{\mu\rho}g_{\nu\sigma} + g_{\mu\sigma}g_{\rho\nu})\ I_2^R ,  \notag \\
& & I_{0\mu\nu}^R = \frac{1}{4} g_{\mu\nu} \ I_0^R, \quad
I_{0\mu\nu\rho\sigma }^R = \frac{1}{24} (g_{\mu\nu}g_{\rho\sigma} +
g_{\mu\rho}g_{\nu\sigma} + g_{\mu\sigma}g_{\rho\nu})\ I_0^R .
\end{eqnarray}
the resulting loop corrections are gauge invariant. Here the superscript "R"
denotes the regularized ILIs.

Note that the introduction of the irreducible loop integrals (ILIs) is
crucial in the loop regularization\cite{wu1, wu2}, and it has been shown
that all Feynman loop integrals can be expressed in terms of the ILIs. In
the definition of ILIs, one important feature is that there should be no
factor of $k^{2}$ in the numerator of loop integration and all the ILIs can
be classified into the scalar type ILIs with the following loop integrand

\begin{equation*}
\frac{1}{(k^2 - M^2)^{\alpha}}
\end{equation*}
and the tensor type ILIs with the following loop integration

\begin{equation*}
\frac{k_{\mu}k_{\nu} \cdots k_{\rho}}{(k^2 - M^2)^{\alpha}}
\end{equation*}

In manipulating the Feynman loop integrals into ILIs, one should always
perform the Dirac algebra and Lorentz index-contraction first to obtain the
ILIs defined by the above "simplest" forms for the one loop case. For
example, the integrand
\begin{equation*}
g^{\mu \nu }\cdot k_{\mu }k_{\nu }/(k^{2}-M^{2})^{2}
\end{equation*}%
should not be written as
\begin{equation*}
g^{\mu \nu }\cdot I_{2\mu \nu }.
\end{equation*}%
Instead it should be expressed as
\begin{equation*}
k^{2}/(k^{2}-M^{2})^{2}
\end{equation*}%
and then rewrite the $k^{2}$ in the numerator in the form $%
(k^{2}-M^{2})+M^{2}$ so as to cancel out the first term by the denominator
to get,
\begin{equation*}
g^{\mu \nu }\cdot k_{\mu }k_{\nu }/(k^{2}-M^{2})^{2}=I_{2}+M^{2}\cdot I_{0}.
\end{equation*}

A simple regularization prescription for the ILIs was realized to yield the
above consistency conditions. The procedure is: rotating to the four
dimensional Euclidean space of momentum, replacing the loop integrating
variable $k^{2}$ and the loop integrating measure $\int {d^{4}k}$ in the
ILIs by the corresponding regularized ones $[k^{2}]_{l}$ and $\int
[d^{4}k]_{l}$:
\begin{eqnarray}
&&\quad k^{2}\rightarrow \lbrack k^{2}]_{l}\equiv k^{2}+M_{l}^{2}\ ,  \notag
\\
&&\int {d^{4}k}\rightarrow \int [d^{4}k]_{l}\equiv
\lim_{N,M_{l}^{2}}\sum_{l=0}^{N}c_{l}^{N}\int {d^{4}k}
\end{eqnarray}%
where $M_{l}^{2}$ ($l=0,1,\ \cdots $) may be regarded as the regulator
masses for the ILIs. The regularized ILIs in the Euclidean space-time are
then given by:
\begin{eqnarray}
I_{-2\alpha }^{R} &=&i(-1)^{\alpha
}\lim_{N,M_{l}^{2}}\sum_{l=0}^{N}c_{l}^{N}\int \frac{d^{4}k}{(2\pi )^{4}}%
\frac{1}{(k^{2}+M^{2}+M_{l}^{2})^{2+\alpha }},  \notag \\
I_{-2\alpha \ \mu \nu }^{R} &=&-i(-1)^{\alpha
}\lim_{N,M_{l}^{2}}\sum_{l=0}^{N}c_{l}^{N}\int \frac{d^{4}k}{(2\pi )^{4}}%
\frac{k_{\mu }k_{\nu }}{(k^{2}+M^{2}+M_{l}^{2})^{3+\alpha }},\hspace{8mm}%
\alpha =-1,0,1,2,...  \notag \\
I_{-2\alpha \ \mu \nu \rho \sigma }^{R} &=&i(-1)^{\alpha
}\lim_{N,M_{l}^{2}}\sum_{l=0}^{N}c_{l}^{N}\int \frac{d^{4}k}{(2\pi )^{4}}%
\frac{k_{\mu }k_{\nu }k_{\rho }k_{\sigma }}{(k^{2}+M^{2}+M_{l}^{2})^{4+%
\alpha }}
\end{eqnarray}%
where the coefficients $c_{l}^{N}$ are chosen to satisfy the following
conditions:
\begin{equation}
\lim_{N,M_{l}^{2}}\sum_{l=0}^{N}c_{l}^{N}(M_{l}^{2})^{n}=0\quad
(n=0,1,\cdots )  \label{cl conditions}
\end{equation}%
with the notation $\lim_{N,M_{l}^{2}}$ denoting the limit $%
\lim_{N,M_{R}^{2}\rightarrow \infty }$. One may take the initial conditions $%
M_{0}^{2}=\mu _{s}^{2}=0$ and $c_{0}^{N}=1$ to recover the original
integrals in the limit $M_{l}^{2}\rightarrow \infty $ ($l=1,2,\cdots $ ).
Such a new regularization is called as Loop Regularization (LORE) \cite{wu1,
wu2}. The prescription in LORE method is very similar to Pauli-Villars
prescription, but two concepts are totally different as the prescription in
the loop regularization is acting on the ILIs rather than on the propagators
as in Pauli-Villars scheme. This is why the Pauli-Villars regularization
violates non-Abelian gauge symmetry, while LORE method can preserve
non-Abelian gauge symmetry.

A simple solution of eq. (\ref{cl conditions}), is to take the string-mode
regulators,
\begin{equation}
M_{l}^{2}=\mu _{s}^{2}+lM_{R}^{2}
\end{equation}%
with $l=1,2,\cdots $, and the coefficients $c_{l}^{N}$ to be of the form,
\begin{equation}
c_{l}^{N}=(-1)^{l}\frac{N!}{(N-l)!l!}  \label{mus}
\end{equation}%
Here $M_{R}$ may be regarded as a basic mass scale of loop regulator . It
has been shown in \cite{wu2} that the above regularization prescription can
be understood in terms of Schwinger proper time formulation with an
appropriate regulating distribution function.

With the string-mode regulators for $M_{l}^{2}$ and $c_{l}^{N}$ in above
equations, the regularized ILIs $I_{2}^{R}$ and $I_{0}^{R}$ can be
calculated to give\cite{wu1, wu2}:
\begin{eqnarray}
I_{2}^{R} &=&\frac{-i}{16\pi ^{2}}\{M_{c}^{2}-\mu ^{2}[\ln \frac{M_{c}^{2}}{%
\mu ^{2}}-\gamma _{w}+1+y_{2}(\frac{\mu ^{2}}{M_{c}^{2}})]\}  \notag
\label{Re} \\
I_{0}^{R} &=&\frac{i}{16\pi ^{2}}[\ln \frac{M_{c}^{2}}{\mu ^{2}}-\gamma
_{w}+y_{0}(\frac{\mu ^{2}}{M_{c}^{2}})]
\end{eqnarray}%
with $\mu ^{2}=\mu _{s}^{2}+M^{2}$, and
\begin{eqnarray}
&&\gamma _{w}\equiv \lim_{N}\{\ \sum_{l=1}^{N}c_{l}^{N}\ln l+\ln [\
\sum_{l=1}^{N}c_{l}^{N}\ l\ln l\ ]\}=\gamma _{E}=0.5772\cdots ,  \notag
\label{y-function} \\
&&y_{0}(x)=\int_{0}^{x}d\sigma \frac{1-e^{-\sigma }}{\sigma },\quad y_{1}(x)=%
\frac{e^{-x}-1+x}{x}  \notag \\
&&y_{2}(x)=y_{0}(x)-y_{1}(x),\quad \lim_{x\rightarrow 0}y_{i}(x)\rightarrow
0,\ i=0,1,2 \\
&&M_{c}^{2}\equiv \lim_{N,M_{R}}M_{R}^{2}\sum_{l=1}^{N}c_{l}^{N}(l\ln
l)=\lim_{N,M_{R}}M_{R}^{2}/\ln N  \notag
\end{eqnarray}%
This indicates that the $\mu _{s}$ sets an IR `cutoff' at $M^{2}=0$ and $%
M_{c}$ provides an UV `cutoff'. For renormalizable quantum field theories, $%
M_{c}$ can be taken to infinity $(M_{c}\rightarrow \infty )$. In a theory
without infrared divergence, $\mu _{s}$ can safely run to $\mu _{s}=0$.
Actually, in the case that $M_{c}\rightarrow \infty $ and $\mu _{s}=0$, one
recovers the initial integral. Also once $M_{R}$ and $N$ are taken to be
infinity, the regularized theory becomes independent of the regularization
prescription. Note that to evaluate the ILIs, products of $\gamma $ matrices
involving loop momentum $k\hspace{-0.17cm}\slash$ such as $k\hspace{-0.17cm}%
\slash\gamma _{\mu }k\hspace{-0.17cm}\slash$ should be reduced to one of the
independent components: $\gamma _{\mu }$, $\sigma _{\mu \nu }$, $\gamma
_{5}\gamma _{\mu }$, $\gamma _{5}$ without $k.$

\section{Massless QED Vacuum Polarization at Two Loop Order}

In this section, we will present our results of vacuum polarization in the
massless Quantum Electrodynamics (QED) at two-loop order. The motivation for
this computation is two-folded: (1) to show how to apply Loop Regularization
(LORE) method to tensor-type two-loop integrals; (2) to give the first
explicit example of applying LORE to the realistic model. We will show that
LORE can preserve Ward Identity and in turn the gauge invariance of QED at
two-loop order. Also, our result reproduces the results in the standard
textbooks like \cite{bjor,Peskin:1995ev,Itzykson:1980rh}. It is an essential
step towards applying LORE to the general gauge theory calculations, such as
the Standard Model of particle physics.

The Lagrangian of the massless QED is:
\begin{equation}
\mathcal{L}_{QED}=-\frac{1}{4}(F_{\mu \nu })^{2}+i\bar{\psi}D\hspace{-0.17cm}%
\slash\psi
\end{equation}%
where, $D_{\mu }=\partial _{\mu }+ieA_{\mu }$ is the covariant derivative
and $F_{\mu \nu }=\partial _{\mu }A_{\nu }-\partial _{\nu }A_{\mu }$ is the
field strength tensor of electromagnetic field. The detailed Feynman rules
for QED are referred to the standard texts, such as \cite%
{bjor,Peskin:1995ev,Itzykson:1980rh}.


\subsection{Regularization and Renormalization of QED at One-Loop Level}

\label{l1}

\begin{figure}[th]
\begin{center}
\includegraphics[width=15cm,clip=true,keepaspectratio=true]{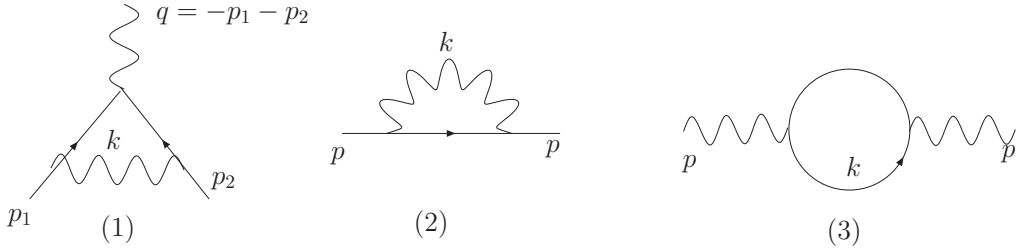}\\[0pt]
\end{center}
\caption{1-Loop 1PI diagrams: Left: Vertex correction; Middle: Electron self
energy; Right: Vacuum Polarization }
\label{qed1}
\end{figure}

In this section, we present the regularization and renormalization for all
the divergent one-particle-irreducible (1PI) diagrams in the massless QED at
one-loop level with the LORE for use in the later calculation. At one-loop
level, the massless QED has three divergent 1PI diagrams requiring
regularization and renormalization. They are shown in Fig.(\ref{qed1})

Using the standard method of Feynman parameters, we get for the vertex
correction
\begin{eqnarray}
-i\Gamma ^{\mu (1)} &=&(-ie)^{3}\int \frac{d^{4}k}{(2\pi )^{4}}(\gamma
^{\sigma }\frac{i}{k\hspace{-0.17cm}\slash+p\hspace{-0.17cm}\slash_{2}}%
\gamma ^{\mu }\frac{i}{k\hspace{-0.17cm}\slash-p\hspace{-0.17cm}\slash_{1}}%
\gamma ^{\rho })\frac{-ig_{\rho \sigma }}{k^{2}}  \notag \\
&=&2\Gamma (3)e^{3}\int dxdy\int \frac{d^{4}k}{(2\pi )^{4}}\frac{[k\hspace{%
-0.17cm}\slash-(1-x)p\hspace{-0.17cm}\slash_{1}-yp\hspace{-0.17cm}\slash%
_{2}]\gamma ^{\mu }[k\hspace{-0.17cm}\slash+xp\hspace{-0.17cm}\slash%
_{1}+(1-y)p\hspace{-0.17cm}\slash_{2}]}{[k^{2}-M_{p_{1},p_{2}}^{2}]^{3}}
\end{eqnarray}%
where $M_{p_{1},p_{2}}^{2}=(xp_{1}-yp_{2})^{2}-xp_{1}^{2}-yp_{2}^{2}$. We
can write this in terms of 1-folded ILIs as,
\begin{eqnarray}
-i\Gamma ^{\mu (1)} &=&2\Gamma (3)e^{3}\int dxdy\{(\gamma ^{\alpha }\gamma
^{\mu }\gamma ^{\beta })I_{0\alpha \beta }^{R}-[(1-x)p\hspace{-0.17cm}\slash%
_{1}+yp\hspace{-0.17cm}\slash_{2}]\gamma ^{\mu }[xp\hspace{-0.17cm}\slash%
_{1}+(1-y)p\hspace{-0.17cm}\slash_{2}]I_{-2}\}  \notag \\
&=&2\Gamma (3)e^{3}\int dxdy\{-\frac{1}{2}\gamma ^{\mu }I_{0}^{R}-[(1-x)p%
\hspace{-0.17cm}\slash_{1}+yp\hspace{-0.17cm}\slash_{2}]\gamma ^{\mu }[xp%
\hspace{-0.17cm}\slash_{1}+(1-y)p\hspace{-0.17cm}\slash_{2}]I_{-2},\}
\end{eqnarray}%
where in the second line we have applied the consistency condition Eq.(\ref%
{cc}) to deal with the regularized logarithmic divergent integral $I_{0\mu
\nu }^{R}$. By using the explicit expression of $I_{0}^{R}$ Eq.(\ref{Re})
and integrating out $I_{-2}$, we can write our result of the one-loop vertex
correction as,
\begin{equation}
-i\Gamma ^{\mu (1)}=-\frac{ie^{3}}{8\pi ^{2}}\int dxdy\{\gamma ^{\mu }[\ln
\frac{M_{c}^{2}}{\mu _{1}^{2}}-\gamma _{\omega }+y_{0}(\frac{\mu _{1}^{2}}{%
M_{c}^{2}})]-\frac{[(1-x)p\hspace{-0.17cm}\slash_{1}+yp\hspace{-0.17cm}\slash%
_{2}]\gamma ^{\mu }[xp\hspace{-0.17cm}\slash_{1}+(1-y)p\hspace{-0.17cm}\slash%
_{2}]}{(xp_{1}-yp_{2})^{2}-xp_{1}^{2}-yp_{2}^{2}}\},  \notag
\end{equation}%
where $\mu _{1}^{2}=\mu _{s}^{2}+M_{p_{1},p_{2}}^{2}$. According to our
renormalization scheme in LORE method discussed in our first paper\cite%
{Huang}, we should define the one-loop vertex counterterm as:
\begin{equation}
-ie\gamma ^{\mu }\delta _{1}=-\frac{ie^{3}}{8\pi ^{2}}\gamma ^{\mu }\int
dxdy(\ln \frac{M_{c}^{2}}{\mu ^{2}}-\gamma _{\omega })=-\frac{ie^{3}}{16\pi
^{2}}\gamma ^{\mu }(\ln \frac{M_{c}^{2}}{\mu ^{2}}-\gamma _{\omega }),
\notag
\end{equation}%
where $\mu $ is introduced as the renormalization scale. With such
definition of $\delta _{1}$, the renormalized vertex correction at one-loop
level is:
\begin{equation*}
-i\Gamma _{R}^{\mu (1)}=-\frac{ie^{3}}{8\pi ^{2}}\int dxdy\{\gamma ^{\mu
}\ln \frac{\mu ^{2}}{\mu _{1}^{2}}-\frac{[(1-x)p\hspace{-0.17cm}\slash_{1}+yp%
\hspace{-0.17cm}\slash_{2}]\gamma ^{\mu }[xp\hspace{-0.17cm}\slash_{1}+(1-y)p%
\hspace{-0.17cm}\slash_{2}]}{(xp_{1}-yp_{2})^{2}-xp_{1}^{2}-yp_{2}^{2}}\},
\end{equation*}%
where we have taken the limit $M_{c}^{2}\rightarrow \infty $ so that $y_{0}(%
\frac{\mu _{1}^{2}}{M_{c}^{2}})\rightarrow 0$ to simplify our final result.

For the electron self-energy diagram, we can use the same procedure while
the calculation is more straightforward,
\begin{eqnarray}
i\Sigma \hspace{-0.17cm}\slash(p) &=&(-ie)^{2}\int \frac{d^{4}k}{(2\pi )^{4}}%
(\gamma ^{\mu }\frac{i}{k\hspace{-0.17cm}\slash+p\hspace{-0.17cm}\slash}%
\gamma ^{\nu })\frac{-ig_{\mu \nu }}{k^{2}}  \notag \\
&=&2e^{2}\int_{0}^{1}dx\int \frac{d^{4}k}{(2\pi )^{4}}\frac{(1-x)p\hspace{%
-0.17cm}\slash}{[k^{2}+x(1-x)p^{2}]^{2}},
\end{eqnarray}%
Using the general formula of LORE, we can regularize the above logarithmic
divergent integral as follows:
\begin{eqnarray}
i\Sigma \hspace{-0.17cm}\slash(p) &=&2e^{2}p\hspace{-0.17cm}\slash%
\int_{0}^{1}dx(1-x)I_{0}^{R}  \notag \\
&=&\frac{ie^{2}}{8\pi ^{2}}p\hspace{-0.17cm}\slash\int_{0}^{1}dx(1-x)[\ln
\frac{M_{c}^{2}}{\mu _{s}^{2}-x(1-x)p^{2}}-\gamma _{\omega }+y_{0}(\frac{\mu
_{s}^{2}-x(1-x)p^{2}}{M_{c}^{2}})].
\end{eqnarray}%
If we introduce the counterterm for the self-energy as:
\begin{equation*}
ip\hspace{-0.17cm}\slash\delta _{2}=-\frac{ie^{2}}{16\pi ^{2}}p\hspace{%
-0.17cm}\slash(\ln \frac{M_{c}^{2}}{\mu }-\gamma _{\omega }),
\end{equation*}%
then the renormalized self-energy for electron is:
\begin{equation}
i\Sigma \hspace{-0.17cm}\slash_{R}(p)=\frac{ie^{2}}{8\pi ^{2}}p\hspace{%
-0.17cm}\slash\int_{0}^{1}dx(1-x)\ln \frac{\mu ^{2}}{\mu _{s}^{2}-x(1-x)p^{2}%
},
\end{equation}%
Note that the counterterm coefficients defined above have the interesting
relationship $\delta _{1}=\delta _{2}$, which is the result of Ward identity
in QED. This relationship is crucial to guarantee the consistency of our
later discussion of renormalization at two-loop level.

Finally, the one-loop vacuum polarization diagram can be written as:
\begin{eqnarray}\label{1-loop vp}
\mathcal{M}^{(1)} &=&(-1)(-ie)^{2}\int \frac{d^{4}k}{(2\pi )^{4}}tr(\gamma
^{\mu }\frac{i}{k\hspace{-0.17cm}\slash}\gamma ^{\nu }\frac{i}{k\hspace{%
-0.17cm}\slash+p\hspace{-0.17cm}\slash})  \notag \\
&=&-e^{2}tr(\gamma ^{\mu }\gamma ^{\alpha }\gamma ^{\nu }\gamma ^{\beta
})\int \frac{d^{4}k}{(2\pi )^{4}}\int_{0}^{1}dx\frac{k_{\alpha }k_{\beta
}-x(1-x)p_{\alpha }p_{\beta }}{[k^{2}+x(1-x)p^{2}]^{2}}
\end{eqnarray}%
By computing the trace of gamma matrices, we obtain the following expression
in terms of ILIs:
\begin{equation}
\mathcal{M}^{(1)}=-4e^{2}\int_{0}^{1}dx\{(2I_{2}^{\mu \nu R}-g^{\mu \nu
}I_{2}^{R})-2x(1-x)(p^{\mu }p^{\nu }-g^{\mu \nu }p^{2})I_{0}^{R}\}.  \notag
\end{equation}%
With help of consistency condition Eq.(\ref{cc}), it is seen that the
quadratically divergent terms in the first parenthesis cancel each other
exactly, and the remaining term is proportional to $(p^{\mu }p^{\nu }-g^{\mu
\nu }p^{2})$, as required by gauge invariance. Using the formula for $%
I_{0}^{R}$, we get,
\begin{equation*}
\mathcal{M}^{(1)}=\frac{ie^{2}}{2\pi ^{2}}(p^{\mu }p^{\nu }-g^{\mu \nu
}p^{2})\int_{0}^{1}dxx(1-x)[\ln \frac{M_{c}^{2}}{\mu _{s}^{2}-x(1-x)p^{2}}%
-\gamma _{\omega }+y_{0}(\frac{\mu _{s}-x(1-x)p^{2}}{M_{c}^{2}})].
\end{equation*}%
Again, if we choose the counterterm for the photon vacuum polarization as:
\begin{eqnarray}\label{1-loop vp}
i(p^{\mu }p^{\nu }-g^{\mu \nu }p^{2})\delta _{3} &=&\frac{ie^{2}}{2\pi ^{2}}%
(p^{\mu }p^{\nu }-g^{\mu \nu }p^{2})\int_{0}^{1}dx\;x(1-x)(\ln \frac{%
M_{c}^{2}}{\mu ^{2}}-\gamma _{\omega })  \notag \\
&=&\frac{ie^{2}}{12\pi ^{2}}(p^{\mu }p^{\nu }-g^{\mu \nu }p^{2})(\ln \frac{%
M_{c}^{2}}{\mu ^{2}}-\gamma _{\omega })
\end{eqnarray}%
the renormalized photon vacuum polarization at one-loop level is:
\begin{equation*}
\mathcal{M}^{(1)}=\frac{ie^{2}}{2\pi ^{2}}(p^{\mu }p^{\nu }-g^{\mu \nu
}p^{2})\int_{0}^{1}dxx(1-x)\ln \frac{\mu ^{2}}{\mu _{s}^{2}-x(1-x)p^{2}},
\end{equation*}%
where we have taken $y_{0}(\frac{\mu _{s}^{2}-x(1-x)p^{2}}{M_{c}^{2}}%
)\rightarrow 0$ when $M_{c}^{2}\rightarrow \infty $ as usual. This completes
our discussion of the regularization and renormalization of massless QED at
one-loop level.


\subsection{Self-Energy Insertion Diagrams}

\label{SEID} Now we compute the photon vacuum polarization at two-loop
order. In this subsection, we calculate the diagrams $(a_{1})$ and $(a_{2})$%
. It is easy to see that the two diagrams are equal and can be recognized as
the insertion of one-loop electron self-energy into the one-loop photon
vacuum polarization.
\begin{figure}[th]
\begin{center}
\includegraphics[scale=0.8]{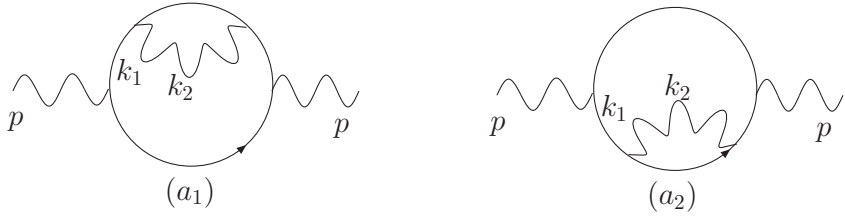}
\end{center}
\caption{Self-Energy Insertion Diagrams. Left: $(a_{1})$; Right: $(a_{2})$}
\label{qedA}
\end{figure}

We first write 1-loop electron self-energy using the UVDP parametrization,
rather than the Feynman parametrization like we did in Sec.\ref{l1}. The
reason lies in the fact that this time the one-loop electron self-energy
appears as the subdiagram in $(a_{1})$ and may contribute to the overlapping
divergence in the following calculation,%
\begin{eqnarray}
i\Sigma \hspace{-0.17cm}\slash^{(1)}(k_{1}) &=&(-ie)^{2}\int \frac{d^{4}k_{2}%
}{(2\pi )^{4}}(\gamma ^{\mu }\frac{i}{k\hspace{-0.17cm}\slash_{1}+k\hspace{%
-0.17cm}\slash_{2}}\gamma ^{\nu })\frac{-ig_{\mu \nu }}{k_{2}^{2}}  \notag
\label{1-loop se} \\
&=&2e^{2}\int_{0}^{\infty }\frac{duu}{(u+1)^{3}}\int \frac{d^{4}k_{2}}{(2\pi
)^{4}}\frac{k\hspace{-0.17cm}\slash_{1}}{[k_{2}^{2}+\frac{u}{(1+u)^{2}}%
k_{1}^{2}]^{2}},
\end{eqnarray}%
By inserting the above result into the expression of one-loop photon vacuum
polarization diagram, we obtain:
\begin{eqnarray}
\mathcal{M}^{(a_{1})} &=&(-ie)^{2}\int \frac{d^{4}k_{1}}{(2\pi )^{4}}%
(-1)tr[\gamma ^{\mu }\frac{i}{k\hspace{-0.17cm}\slash_{1}}i\Sigma \hspace{%
-0.17cm}\slash^{(1)}(k_{1})\frac{i}{k\hspace{-0.17cm}\slash_{1}}\gamma ^{\nu
}\frac{i}{k\hspace{-0.17cm}\slash_{1}+p\hspace{-0.17cm}\slash}]  \notag \\
&=&-2ie^{4}\int_{0}^{\infty }\frac{du(u+1)}{u}\int \frac{d^{4}k_{1}}{(2\pi
)^{4}}\int \frac{d^{4}k_{2}}{(2\pi )^{4}}tr(\gamma ^{\mu }\gamma ^{\rho
}\gamma ^{\nu }\gamma ^{\sigma })\int_{0}^{1}dx  \notag \\
&&\frac{k_{1\rho }(k_{1}+p)_{\sigma }}{[\frac{(1+u)^{2}}{u}%
k_{2}^{2}+k_{1}^{2}]^{2}[k_{1}^{2}+2xk_{1}p+xp^{2}]^{2}}.
\end{eqnarray}%
Note that we have used the usual Feynman parameter $x$ to combine the
denominators involving $k_{1}$ only, rather than combine all of the three
factors once at all. This transforms the Feynman integral $\mathcal{M}%
^{(a_{1})}$ into the $\alpha \beta \gamma -$like integrals (we have already
combine two factors in Eq.(\ref{1-loop se}) with UVDP parameter $u$ ). The
advantage of this procedure is that it effectively separates the UV
divergences from the IR ones in the parameter space. To see this more
explicitly, our first paper\cite{Huang} already explicitly showed that the $%
\alpha \beta \gamma $ integrals contain the most general UV divergence
structure. In other words, the UV divergence can only be contained in $%
\alpha \beta \gamma $ integrals and their corresponding UVDP parameter
integrals, rather in the Feynman parameter space. If we find some divergence
in integrating Feynman parameter $x$, then it must be the IR divergence,
rather than the UV one. Since we have already written the integral into a
generalized $\alpha \beta \gamma $ form, for the rest factors, we will apply
the UVDP parameters to combine them,
\begin{eqnarray}
\mathcal{M}^{(a_{1})} &=&-8ie^{4}\Gamma (4)\int_{0}^{\infty }\frac{du(1+u)}{u%
}\int_{0}^{1}dx\int_{0}^{\infty }\frac{dv_{1}v_{1}}{(1+v_{1})^{4}}\int \frac{%
d^{4}k_{1}}{(2\pi )^{4}}\int \frac{d^{4}k_{2}}{(2\pi )^{4}}  \notag \\
&&\frac{(2k_{1}^{\mu }k_{1}^{\nu }-g^{\mu \nu }k_{1}^{2})-\frac{xv_{1}}{%
1+v_{1}}(1-\frac{xv_{1}}{1+v_{1}})(2p^{\mu }p^{\nu }-g^{\mu \nu }p^{2})}{%
[k_{1}^{2}+\frac{xv_{1}}{1+v_{1}}(1-\frac{xv_{1}}{1+v_{1}})p^{2}+\frac{%
(1+u)^{2}}{u(1+v_{1})}k_{2}^{2}]^{4}},
\end{eqnarray}

Note also that, in the above calculation, we used the traditional technique
of completing squares for each internal loop momentum and then shifting the
origin of it in order to eliminate the terms linear in the momenta. The
resulting formula appears less symmetric in the UVDP variables than that
with the general formulae in Ref.\cite{Huang}. However, this traditional
method is more flexible and more natural in practice when the numerator is a
complicated function of internal and external momenta like the present case,
rather than a constant for the scalar integrals. In the following
calculation of Fig. (\ref{qedB}), we will also use this traditional method.

Since there is no cross terms between $k_{1}$ and $k_{2}$ in the
denominator, we can easily integrate out $k_{1}$ and $k_{2}$ with the LORE
method sequentially:
\begin{eqnarray}
\mathcal{M}^{(a_{1})} &=&\frac{8ie^{4}\Gamma (4)}{16\pi ^{2}}%
\int_{0}^{\infty }\frac{du(1+u)}{u}\int_{0}^{1}dx\int_{0}^{\infty }\frac{%
dv_{1}v_{1}}{(1+v_{1})^{4}}\{\frac{1}{6}g^{\mu \nu }\frac{u(1+v_{1})}{%
(1+u)^{2}}I_{2}^{R}(\mu ^{2})  \notag  \label{aa} \\
&&-\frac{1}{6}(2p^{\mu }p^{\nu }-g^{\mu \nu }p^{2})\frac{xv_{1}}{1+v_{1}}(1-%
\frac{xv_{1}}{1+v_{1}})\frac{u^{2}(1+v_{1})^{2}}{(1+u)^{4}}I_{0}^{R}(\mu
^{2})\}  \notag \\
&=&\mathcal{M}_{2}^{(a_{1})}+\mathcal{M}_{0}^{(a_{1})},
\end{eqnarray}%
where $\mu ^{2}\equiv \mu _{s}^{2}-\frac{uxv_{1}}{(1+u)^{2}}(1-\frac{xv_{1}}{%
1+v_{1}})p^{2}$ and $\mathcal{M}_{2(0)}^{(a_{1})}$ denotes quadratic
(logarithmic) divergent part from the integration of $k_{2}$.

Note that since our integrals in the present paper do not contain any IR
divergences, we can set $\mu _{s}^{2}\rightarrow 0$ which also plays the
role of IR regulator in the LORE method. Moreover, because the $y_{i}(\frac{%
\mu ^{2}}{M_{c}^{2}})\rightarrow 0$ in the limit of $\frac{\mu ^{2}}{%
M_{c}^{2}}\rightarrow 0$, so these functions do not contribute to our final
results and can be ignored.

To make our discussion more concise, we only present our final regulated
result for the diagram $(a_{1})$, while relegating our calculational details
of the UVDP parameter integrations to the Appendix \ref{AUVDP}.
\begin{eqnarray}
\mathcal{M}^{(a_{1})} &\sim &\frac{4ie^{4}}{(16\pi ^{2})^{2}}\{g^{\mu \nu
}M_{c}^{2}(\ln \frac{M_{c}^{2}}{-q_{o}^{2}}-\gamma _{\omega })+\frac{1}{6}%
g^{\mu \nu }p^{2}(\ln \frac{M_{c}^{2}}{-q_{o}^{2}}-\gamma _{\omega })  \notag
\label{af} \\
&&+(g^{\mu \nu }p^{2}-p^{\mu }p^{\nu })\cdot \lbrack \frac{1}{3}(\ln \frac{%
M_{c}^{2}}{-p^{2}}-\gamma _{\omega })(\ln \frac{M_{c}^{2}}{-q_{o}^{2}}%
-\gamma _{\omega })-\frac{1}{6}(\ln \frac{M_{c}^{2}}{-q_{o}^{2}}-\gamma
_{\omega })^{2}+\frac{1}{6}\alpha _{\omega }  \notag \\
&&-\frac{5}{18}(\ln \frac{M_{c}^{2}}{-p^{2}}-\gamma _{\omega })+\frac{11}{9}%
(\ln \frac{M_{c}^{2}}{-q_{o}^{2}}-\gamma _{\omega })]\}
\end{eqnarray}

\begin{figure}[th]
\begin{center}
\includegraphics[scale=0.6]{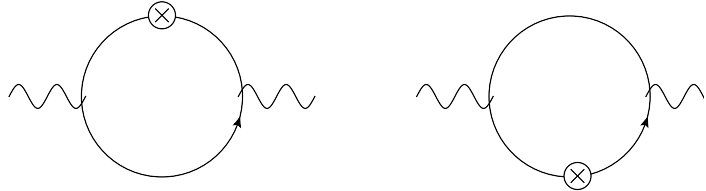}
\end{center}
\caption{Counterterm Insertion Diagrams for Diagram Group $(a)$. Left: $%
(a_{1}^{\prime })$; Right: $(a_{2}^{\prime })$}
\label{Ac}
\end{figure}

\begin{figure}[th]
\begin{center}
\includegraphics[width=8cm,clip=true,keepaspectratio=true]{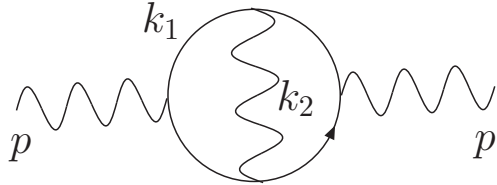}\\[0pt]
\end{center}
\caption{Vertex Correction Insertion Diagram $(b)$}
\label{qedB}
\end{figure}

In order to see the cancellation of harmful or nonlocal terms for diagram $%
(a_{1})$, we need to compute its corresponding counterterm diagram $%
(a_{1}^{\prime })$, as shown on the left in Fig. (\ref{Ac}),
and this diagram contributes
\begin{eqnarray}
\mathcal{M}^{(a_{1}^{\prime })} &=&\mathcal{M}^{(a_{2}^{\prime
})}=(-ie)^{2}(-1)\int \frac{d^{4}k}{(2\pi )^{4}}tr[\gamma ^{\mu }\frac{i}{k%
\hspace{-0.17cm}\slash+p\hspace{-0.17cm}\slash}\gamma ^{\nu }\frac{i}{k%
\hspace{-0.17cm}\slash}(ik\hspace{-0.17cm}\slash\delta _{2})\frac{i}{k%
\hspace{-0.17cm}\slash}]  \notag \\
&=&(-\delta _{2})(-1)(-ie)^{2}\int \frac{d^{4}k}{(2\pi )^{4}}tr(\gamma ^{\mu
}\frac{i}{k\hspace{-0.17cm}\slash+p\hspace{-0.17cm}\slash}\gamma ^{\nu }%
\frac{i}{k\hspace{-0.17cm}\slash})  \notag \\
&=&-\frac{4ie^{4}}{(16\pi ^{2})^{2}}(g^{\mu \nu }p^{2}-p^{\mu }p^{\nu })[%
\frac{1}{3}(\ln \frac{M_{c}^{2}}{\mu ^{2}}-\gamma _{\omega })^{2}  \notag \\
&+&\frac{1}{3}(\ln \frac{M_{c}^{2}}{\mu ^{2}}-\gamma _{\omega })\ln \frac{%
\mu ^{2}}{-p^{2}}+\frac{5}{9}(\ln \frac{M_{c}^{2}}{\mu ^{2}}-\gamma _{\omega
})]
\end{eqnarray}%
Here we recognized that the internal momentum integration is essentially the
same as the photon vacuum polarization at one-loop order and the result Eq.(%
\ref{1-loop vp}) can be directly applied.

By summing up the diagrams $(a_{1})$ and $(a_{1}^{\prime })$, we can obtain:
\begin{eqnarray}
\mathcal{M}^{(a_{1}+a_{1}^{\prime })} &=&\frac{4ie^{4}}{(16\pi ^{2})^{2}}%
\{g^{\mu \nu }M_{c}^{2}(\ln \frac{M_{c}^{2}}{-q_{o}^{2}}-\gamma _{\omega
})+g^{\mu \nu }p^{2}[\frac{1}{6}(\ln \frac{M_{c}^{2}}{-q_{o}^{2}}-\gamma
_{\omega })-\frac{5}{36}]  \notag  \label{as} \\
&&+(g^{\mu \nu }p^{2}-p^{\mu }p^{\nu })[-\frac{1}{6}(\ln \frac{M_{c}^{2}}{%
\mu ^{2}}-\gamma _{\omega })^{2}+\frac{7}{18}(\ln \frac{M_{c}^{2}}{\mu ^{2}}%
-\gamma _{\omega })+\frac{1}{3}\ln \frac{\mu ^{2}}{-p^{2}}\ln \frac{\mu ^{2}%
}{-q_{o}^{2}}  \notag \\
&&-\frac{1}{6}\ln ^{2}\frac{\mu ^{2}}{-q_{o}^{2}}-\frac{5}{18}\ln \frac{\mu
^{2}}{-p^{2}}+\frac{11}{9}\ln \frac{\mu ^{2}}{-q_{o}^{2}}-\frac{1}{6}\alpha
_{\omega }-\frac{17}{12}+\frac{\pi ^{2}}{18}]\}.
\end{eqnarray}%
Note that there is still a harmful divergence $\frac{4ie^{4}}{(16\pi
^{2})^{2}}g^{\mu \nu }M_{c}^{2}\ln (-q_{o}^{2})$, which also breaks the
underlying gauge invariance and its associated Ward identity. From the
general argument of gauge invariance, we expect that terms like this should
not appear in the final result of the two-loop vacuum polarization. As we
will see in the following, the result of diagram (b) contains same term with
the opposite sign. Therefore, the addition of $(a_{1})$, $(a_{2})$ and $(b)$
will eliminate this unwanted term, and recover the gauge invariance and
locality of the underlying field theory.


\subsection{Vertex Correction Insertion Diagram}

\indent Now we compute the more challenging diagram in Fig 4. Following
the standard Feynman rules of QED, we can write down the explicit expression
$\mathcal{M}^{(b)}$
\begin{eqnarray}
\mathcal{M}^{(b)} &=&(-ie)^{4}\int \frac{d^{4}k_{2}}{(2\pi )^{4}}\int \frac{%
d^{4}k_{1}}{(2\pi )^{4}}(-1)tr(\gamma ^{\mu }\frac{i}{k\hspace{-0.17cm}\slash%
_{1}}\gamma ^{\rho }\frac{i}{k\hspace{-0.17cm}\slash_{1}+k\hspace{-0.17cm}%
\slash_{2}}\gamma ^{\nu }\frac{i}{k\hspace{-0.17cm}\slash_{1}+k\hspace{%
-0.17cm}\slash_{2}+p\hspace{-0.17cm}\slash}\gamma ^{\sigma }\frac{i}{k%
\hspace{-0.17cm}\slash_{1}+p\hspace{-0.17cm}\slash})\frac{-ig_{\mu \nu }}{%
k_{2}^{2}}.  \notag  \label{Borig} \\
&=&ie^{4}\int \frac{d^{4}k_{2}}{(2\pi )^{4}}\int \frac{d^{4}k_{1}}{(2\pi
)^{4}}\int_{0}^{1}dx_{1}\int_{0}^{1}dx_{2}  \notag \\
&&\frac{tr[\gamma ^{\mu }(k\hspace{-0.17cm}\slash_{1}-x_{1}p\hspace{-0.17cm}%
\slash)\gamma ^{\rho }(k\hspace{-0.17cm}\slash_{1}+k\hspace{-0.17cm}\slash%
_{2}-x_{1}p\hspace{-0.17cm}\slash)\gamma ^{\nu }(k\hspace{-0.17cm}\slash%
_{1}+k\hspace{-0.17cm}\slash_{2}+(1-x_{1})p\hspace{-0.17cm}\slash)\gamma
_{\rho }(k\hspace{-0.17cm}\slash_{1}+(1-x_{1})p\hspace{-0.17cm}\slash)]}{%
(k_{1}^{2}+x_{1}(1-x_{1})p^{2})^{2}[(k_{1}+k_{2}+(x_{2}-x_{1})p)^{2}+x_{2}(1-x_{2})p^{2}]^{2}k_{2}^{2}%
}
\end{eqnarray}%
where we have used Feynman parameters to combine some of the denominators.
Next, in order to integrate over momentum $k_{1}$, we apply the UVDP
parametrization to combine the factors involving $k_{1}$ in the denominator
to get
\begin{eqnarray}
&&\frac{1}{%
(k_{1}^{2}+x_{1}(1-x_{1})p^{2})^{2}[(k_{1}+k_{2}+(x_{2}-x_{1})p)^{2}+x_{2}(1-x_{2})p^{2}]^{2}k_{2}^{2}%
}  \notag  \label{combination} \\
&=&\Gamma (4)\int_{0}^{\infty }\prod_{i=i}^{2}\frac{dv_{i}}{(1+v_{i})^{2}}%
\frac{\delta (1-\sum_{i=1}^{2}\frac{1}{1+v_{i}})\frac{1}{(1+v_{1})(1+v_{2})}%
}{\mathcal{D}^{4}k_{2}^{2}},
\end{eqnarray}%
where we define
\begin{eqnarray}
\mathcal{D} &\equiv &\frac{1}{1+v_{1}}[k_{1}^{2}+x_{1}(1-x_{1})p^{2}]+\frac{1%
}{1+v_{2}}\{[k_{1}+k_{2}+(x_{2}-x_{1})p]^{2}+x_{2}(1-x_{2})p^{2}\}  \notag \\
&=&(k_{1}+\frac{1}{1+v_{2}}k_{2}+\frac{x_{2}-x_{1}}{1+v_{2}}p)^{2}  \notag \\
&&+\frac{1}{(1+v_{1})(1+v_{2})}[k_{2}+(x_{2}-x_{1})p]^{2}+[\frac{%
x_{1}(1-x_{1})}{1+v_{1}}+\frac{x_{2}(1-x_{2})}{1+v_{2}}]p^{2}  \notag \\
&\equiv &(k_{1}+\frac{1}{1+v_{2}}k_{2}+\frac{x_{2}-x_{1}}{1+v_{2}}%
p)^{2}+M_{k_{2},p}^{2}.
\end{eqnarray}%
Then we make the following translation
\begin{equation}\label{translation}
k_{1}\longrightarrow k_{1}-\frac{1}{1+v_{2}}k_{2}-\frac{x_{2}-x_{1}}{1+v_{2}}%
p.
\end{equation}%
By expanding the trace of the product of gamma matrices and contracting the
Lorentz indices, we write the numerator as,
\begin{eqnarray}
\mathcal{N} &=&tr[\gamma ^{\mu }(k\hspace{-0.17cm}\slash_{1}+l\hspace{-0.17cm%
}\slash_{1})\gamma ^{\rho }(k\hspace{-0.17cm}\slash_{1}+l\hspace{-0.17cm}%
\slash_{2})\gamma ^{\nu }(k\hspace{-0.17cm}\slash_{1}+l\hspace{-0.17cm}\slash%
_{3})\gamma _{\rho }(k\hspace{-0.17cm}\slash_{1}+l\hspace{-0.17cm}\slash%
_{4})]  \notag \\
&=&(-8)\{g^{\mu \nu }k_{1}^{4}+[2k_{1}^{\mu }k_{1}^{\nu }l_{3}\cdot
l_{4}-g^{\mu \nu }k_{1}^{2}l_{3}\cdot l_{4}-2k_{1}^{\mu }l_{4}^{\nu
}l_{3}\cdot k_{1}-2l_{3}^{\mu }k_{1}^{\nu }l_{4}\cdot k_{1}  \notag \\
&&+2g^{\mu \nu }l_{3}\cdot k_{1}~l_{4}\cdot k_{1}+l_{4}^{\mu }l_{3}^{\nu
}k_{1}^{2}+l_{3}^{\mu }l_{4}^{\nu }k_{1}^{2}]+[l_{4}^{\mu }l_{2}^{\nu
}k_{1}^{2}-l_{2}^{\mu }l_{4}^{\nu }k_{1}^{2}+g^{\mu \nu }l_{2}\cdot
l_{4}k_{1}^{2}]  \notag \\
&&+[-2k_{1}^{\mu }k_{1}^{\nu }l_{2}\cdot l_{3}+g^{\mu \nu
}k_{1}^{2}l_{2}\cdot l_{3}+2k_{1}^{\mu }l_{3}^{\nu }l_{2}\cdot
k_{1}+2k_{1}^{\mu }l_{2}^{\nu }l_{3}\cdot k_{1}-l_{3}^{\mu }l_{2}^{\nu
}k_{1}^{2}-l_{2}^{\mu }l_{3}^{\nu }k_{1}^{2}]  \notag \\
&&+[-l_{3}^{\mu }l_{1}^{\nu }k_{1}^{2}+l_{1}^{\mu }l_{3}^{\nu
}k_{1}^{2}+g^{\mu \nu }l_{1}\cdot l_{3}k_{1}^{2}]+[-2k_{1}^{\mu }k_{1}^{\nu
}l_{1}\cdot l_{4}+g^{\mu \nu }k_{1}^{2}l_{1}\cdot l_{4}+2l_{4}^{\mu
}k_{1}^{\nu }l_{1}\cdot k_{1}  \notag \\
&&+2l_{1}^{\mu }k_{1}^{\nu }l_{4}\cdot k_{1}-l_{4}^{\mu }l_{1}^{\nu
}k_{1}^{2}-l_{1}^{\mu }l_{4}^{\mu }k_{1}^{2}]+[2k_{1}^{\mu }k_{1}^{\nu
}l_{1}\cdot l_{2}-g^{\mu \nu }k_{1}^{2}l_{1}\cdot l_{2}-2l_{2}^{\mu
}k_{1}^{\nu }l_{1}\cdot k_{1}  \notag \\
&&-2k_{1}^{\mu }l_{1}^{\nu }l_{2}\cdot k_{1}+2g^{\mu \nu }l_{1}\cdot
k_{1}~l_{2}\cdot k_{1}+l_{1}^{\mu }l_{2}^{\nu }k_{1}^{2}+l_{2}^{\mu
}l_{1}^{\nu }k_{1}^{2}]\}+L_{1234}^{\mu \nu },
\end{eqnarray}%
where $L_{1234}^{\mu \nu }\equiv tr[\gamma ^{\mu }l\hspace{-0.17cm}\slash%
_{1}\gamma ^{\rho }l\hspace{-0.17cm}\slash_{2}\gamma ^{\nu }l\hspace{-0.17cm}%
\slash_{3}\gamma _{\rho }l\hspace{-0.17cm}\slash_{4}]$. Note that in above
expression, we have used $l_{i}$s to simplify our notation which represent
the four factors in the Eq.(\ref{Borig}) after the translation of Eq.(\ref{translation}). Due to their complexity, we do not write them out explicitly.

After the manipulation above, the integration over the loop momentum $k_{1}$
is straightforward,
\begin{eqnarray}
\mathcal{M}^{(b)} &=&\frac{8e^{4}\Gamma (4)}{16\pi ^{2}}\int_{0}^{1}%
\prod_{i=1}^{2}dx_{i}\int_{0}^{\infty }\prod_{i=1}^{2}\frac{dv_{i}}{%
(1+v_{i})^{3}}\delta (1-\sum_{i=1}^{2}\frac{1}{1+v_{i}})\int \frac{d^{4}k_{2}%
}{(2\pi )^{4}}\frac{1}{k_{2}^{2}}\{g^{\mu \nu }(16\pi ^{2}I_{0}^{R}-\frac{5}{%
6})  \notag \\
&&+\frac{1}{3M_{k_{2},p}^{2}}l_{4}^{\mu }l_{3}^{\nu }+\frac{1}{%
3M_{k_{2},p}^{2}}(l_{4}^{\mu }l_{2}^{\nu }-l_{2}^{\mu }l_{4}^{\nu }+g^{\mu
\nu }l_{2}\cdot l_{4})+\frac{1}{6M_{k_{2},p}^{2}}(g^{\mu \nu }l_{2}\cdot
l_{3}-l_{2}^{\mu }l_{3}^{\nu }-l_{3}^{\mu }l_{2}^{\nu })  \notag \\
&&+\frac{1}{3M_{k_{2},p}^{2}}(l_{1}^{\mu }l_{3}^{\nu }-l_{3}^{\mu
}l_{1}^{\nu }+g^{\mu \nu }l_{1}\cdot l_{3})+\frac{1}{6M_{k_{2},p}^{2}}%
(g^{\mu \nu }l_{1}\cdot l_{4}-l_{1}^{\mu }l_{4}^{\nu }-l_{4}^{\mu
}l_{1}^{\nu })  \notag \\
&&+\frac{1}{3M_{k_{2},p}^{2}}l_{1}^{\mu }l_{2}^{\nu }+\frac{1}{%
6M_{k_{2},p}^{4}}L_{1234}^{\mu \nu }\}.
\end{eqnarray}

Now we integrate over the loop momentum $k_{2}$. By using the UVDP
parametrization, we combine the two factors containing $k_{2}^{2}$ in the
denominators
\begin{eqnarray}
\frac{1}{k_{2}^{2}M_{k_{2},p}^{2}} &=&\int_{0}^{\infty }\frac{du}{(1+u)^{2}}%
\frac{1}{[(1-\frac{1}{1+u})k_{2}^{2}+\frac{1}{1+u}M_{k_{2},p}^{2}]^{2}}
\notag \\
&=&\int_{0}^{\infty }du\frac{1}{(uk_{2}^{2}+M_{k_{2},p}^{2})^{2}}  \notag \\
&=&\int_{0}^{\infty }du\frac{1}{[u+\frac{1}{(1+v_{1})(1+v_{2})}]^{2}}\frac{1%
}{[(k_{2}+\frac{x_{2}-x_{1}}{1+u(1+v_{1})(1+v_{2})}p)^{2}-\mu _{u}^{2}]^{2}},
\end{eqnarray}%
\begin{equation*}
\frac{1}{k_{2}^{2}M_{k_{2},p}^{4}}=2\int_{0}^{\infty }du\frac{1}{[u+\frac{1}{%
(1+v_{1})(1+v_{2})}]^{3}}\frac{1}{[(k_{2}+\frac{x_{2}-x_{1}}{%
1+u(1+v_{1})(1+v_{2})}p)^{2}-\mu _{u}^{2}]^{3}},
\end{equation*}%
\begin{eqnarray}
\frac{1}{k_{2}^{2}}I_{0}^{R} &=&\int \frac{[d^{4}k_{1}]_{l}}{(2\pi )^{4}}%
\frac{1}{k_{2}^{2}(k_{1}^{2}-M_{k_{2},p}^{2})^{2}}  \notag  \label{para_u} \\
&=&-\int \frac{[d^{4}k_{1}]_{l}}{(2\pi )^{4}}\frac{\Gamma (3)}{\Gamma
(2)\Gamma (1)}\int_{0}^{\infty }du\frac{1}{%
[k_{1}^{2}-uk_{2}^{2}-M_{k_{2},p}^{2}]^{3}}  \notag \\
&=&\frac{1}{16\pi ^{2}}\int_{0}^{\infty }du\frac{1}{u+\frac{1}{%
(1+v_{1})(1+v_{2})}}\frac{1}{(k_{2}+\frac{x_{2}-x_{1}}{1+u(1+v_{1})(1+v_{2})}%
p)^{2}-\mu _{u}^{2}},
\end{eqnarray}%
where we have defined
\begin{eqnarray}
\mu _{u}^{2} &\equiv &\{\frac{1}{u+\frac{1}{(1+v_{1})(1+v_{2})}}[\frac{%
(x_{2}-x_{1})^{2}}{(1+v_{1})(1+v_{2})}+\frac{x_{1}(1-x_{1})}{1+v_{1}}+\frac{%
x_{2}(1-x_{2})}{1+v_{2}}]  \notag  \label{mu_u} \\
&&-\frac{1}{[u+\frac{1}{(1+v_{1})(1+v_{2})}]^{2}}\frac{(x_{2}-x_{1})^{2}}{%
(1+v_{1})^{2}(1+v_{2})^{2}}\}(-p^{2}).
\end{eqnarray}

After performing Wick rotation of $k_{2}$, we can integrate out $k_{2}$ with
the LORE method, and this naturally separates the $\mathcal{M}^{(b)}$ into
four parts according to the different powers of the factor $\frac{1}{u+\frac{%
1}{(1+v_{1})(1+v_{2})}}$. The resulting expression is lengthy, so in order
to keep the paper in a readable length, we do not write it here. Appendix %
\ref{BUVDP} gives the explicit form for each part, as well as the careful
calculation of the remaining UVDP parameter integrations. We recommend the
interested reader to resort to the Appendix \ref{BUVDP} for such details.
Below we only present our final results for the vertex correction insertion
diagram $(b)$:
\begin{eqnarray}
\mathcal{M}^{(b)} &\sim &-\frac{8ie^{4}}{(16\pi ^{2})^{2}}\{g^{\mu \nu
}M_{c}^{2}(\ln \frac{M_{c}^{2}}{-q_{o}^{2}}-\gamma _{\omega })+\frac{1}{6}%
g^{\mu \nu }p^{2}(\ln \frac{M_{c}^{2}}{-q_{o}^{2}}-\gamma _{\omega })  \notag
\label{bf} \\
&&+(g^{\mu \nu }p^{2}-p^{\mu }p^{\nu })\cdot \lbrack \frac{1}{3}(\ln \frac{%
M_{c}^{2}}{-p^{2}}-\gamma _{\omega })(\ln \frac{M_{c}^{2}}{-q_{o}^{2}}%
-\gamma _{\omega })+\frac{19}{18}(\ln \frac{M_{c}^{2}}{-p^{2}}-\gamma
_{\omega })]  \notag \\
&&-\frac{1}{3}(\ln \frac{M_{c}^{2}}{-q_{o}^{2}}-\gamma _{\omega })+\frac{1}{3%
}[\frac{13}{6}(\ln \frac{M_{c}^{2}}{-q_{o}^{2}(1+V)}-\gamma _{\omega })-%
\frac{1}{2}(\ln \frac{M_{c}^{2}}{-q_{o}^{2}}-\gamma _{\omega })^{2}+\frac{1}{%
2}\alpha _{\omega }]\}.  \notag
\end{eqnarray}

\begin{figure}[ht]
\begin{center}
\includegraphics[scale=0.6]{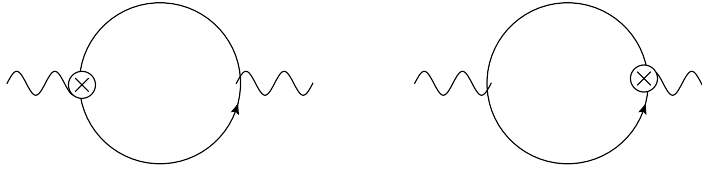}
\end{center}
\caption{Counterterm Insertion Diagrams for Diagram $(b)$. Left: $%
(b_1^\prime)$; Right: $(b_2^\prime)$}
\label{Bc}
\end{figure}

The counterterm diagrams $(b_{1}^{\prime })$ and $(b_{2}^{\prime })$ for $%
(b) $ can be computed directly and the result is
\begin{eqnarray}
\mathcal{M}^{(b_{1}^{\prime })} &=&\mathcal{M}^{(b_{2}^{\prime
})}=(-1)(-ie)^{2}\int \frac{d^{4}k}{(2\pi )^{4}}tr[(\gamma ^{\mu }\delta
_{1})\frac{i}{k\hspace{-0.17cm}\slash+p\hspace{-0.17cm}\slash}\gamma ^{\nu }%
\frac{i}{k\hspace{-0.17cm}\slash}]  \notag \\
&=&\frac{4ie^{4}}{(16\pi ^{2})^{2}}(g^{\mu \nu }p^{2}-p^{\mu }p^{\nu })[%
\frac{1}{3}(\ln \frac{M_{c}^{2}}{\mu ^{2}}-\gamma _{\omega })^{2}+\frac{1}{3}%
(\ln \frac{M_{c}^{2}}{\mu ^{2}}-\gamma _{\omega })\ln \frac{\mu ^{2}}{-p^{2}}%
+\frac{5}{9}(\ln \frac{M_{c}^{2}}{\mu ^{2}}-\gamma _{\omega })],  \notag \\
&&
\end{eqnarray}%
and the summation of $(b)$, $(b_{1}^{\prime })$ and $(b_{2}^{\prime })$
gives:
\begin{eqnarray}
\mathcal{M}^{(b+b_{1}^{\prime }+b_{2}^{\prime })} &\sim &-\frac{8ie^{4}}{%
(16\pi ^{2})^{2}}\{g^{\mu \nu }M_{c}^{2}(\ln \frac{M_{c}^{2}}{-q_{o}^{2}}%
-\gamma _{\omega })+\frac{1}{6}g^{\mu \nu }p^{2}(\ln \frac{M_{c}^{2}}{%
-q_{o}^{2}}-\gamma _{\omega })  \notag  \label{bs} \\
&&+(g^{\mu \nu }p^{2}-p^{\mu }p^{\nu })\cdot \lbrack -\frac{1}{6}(\ln \frac{%
M_{c}^{2}}{\mu ^{2}}-\gamma _{\omega })^{2}+\frac{8}{9}(\ln \frac{M_{c}^{2}}{%
\mu ^{2}}-\gamma _{\omega })+\frac{1}{3}\ln \frac{\mu ^{2}}{-p^{2}}\ln \frac{%
\mu ^{2}}{-q_{o}^{2}}  \notag \\
&&-\frac{1}{6}\ln ^{2}\frac{\mu ^{2}}{-q_{o}^{2}}+\frac{19}{18}\ln \frac{\mu
^{2}}{-p^{2}}+\frac{7}{18}\ln \frac{\mu ^{2}}{-q_{o}^{2}}-\frac{1}{6}\alpha
_{\omega }]\}.
\end{eqnarray}%
As expected from the argument in the end of the subsection \ref{SEID}, the
results of summing over the diagrams $(b+b_{1}^{\prime }+b_{2}^{\prime })$
also possess a harmful divergence which violates the gauge-invariance . As
we will see, this term has the right value to cancel the harmful divergence
in the diagram group $(a)$.


\subsection{The Results}

With the results Eqs.(\ref{as}) and (\ref{bs}) for the diagrams $%
(a_1+a^\prime_1)$ and $(b+b_1^\prime+b_2^\prime)$ at hands, it is
straightforward to obtain the final result of QED vacuum polarization at
two-loop order:
\begin{eqnarray}
\mathcal{M}^{\text{final}} &=& 2\mathcal{M}^{(a_1+a^{\prime }_1)}+\mathcal{M}%
^{(b+b^{\prime }_1+b^{\prime }_2)}  \notag \\
&\sim& \frac{8 i e^4}{(16\pi^2)^2} \{g^{\mu\nu}M_c^2(\ln\frac{M_c^2}{-q_o^2}%
-\gamma_\omega) +g^{\mu\nu}p^2[\frac{1}{6}(\ln\frac{M_c^2}{-q_o^2}%
-\gamma_\omega)-\frac{5}{36}]  \notag \\
&& +(g^{\mu\nu}p^2-p^\mu p^\nu)[-\frac{1}{6}(\ln\frac{M_c^2}{\mu^2}%
-\gamma_\omega)^2 +\frac{7}{18}(\ln\frac{M_c^2}{\mu^2}-\gamma_\omega) +\frac{%
1}{3}\ln\frac{\mu^2}{-p^2}\ln\frac{\mu^2}{-q_o^2}  \notag \\
&&-\frac{1}{6}\ln^2\frac{\mu^2}{-q_o^2} -\frac{5}{18}\ln\frac{\mu^2}{-p^2} +%
\frac{11}{9}\ln\frac{\mu^2}{-q_o^2}-\frac{1}{6}\alpha_\omega-\frac{17}{12}+%
\frac{\pi^2}{18}] \}  \notag \\
&&-\frac{8ie^4}{(16\pi^2)^2}\{g^{\mu\nu}M_c^2(\ln\frac{M_c^2}{-q_o^2}%
-\gamma_\omega) +\frac{1}{6}g^{\mu\nu}p^2(\ln\frac{M_c^2}{-q_o^2}%
-\gamma_\omega)  \notag \\
&& +(g^{\mu\nu}p^2-p^\mu p^\nu)\cdot[-\frac{1}{6}(\ln\frac{M_c^2}{\mu^2}%
-\gamma_\omega)^2 +\frac{8}{9}(\ln\frac{M_c^2}{\mu^2}-\gamma_\omega) +\frac{1%
}{3}\ln\frac{\mu^2}{-p^2}\ln\frac{\mu^2}{-q_o^2}  \notag \\
&& -\frac{1}{6}\ln^2\frac{\mu^2}{-q_o^2}+\frac{19}{18}\ln\frac{\mu^2}{-p^2} +%
\frac{7}{18}\ln\frac{\mu^2}{-q_o^2}-\frac{1}{6}\alpha_\omega]\}  \notag \\
&\sim& \frac{8ie^4}{(16\pi^2)^2}(g^{\mu\nu}p^2-p^\mu p^\nu)[-\frac{4}{3}(\ln%
\frac{M_c^2}{-p^2}-\gamma_\omega)+\frac{5}{6}(\ln\frac{M_c^2}{-q_o^2}%
-\gamma_\omega)].
\end{eqnarray}

Note that there is a free mass scale $-q_{o}^{2}$ in the above expression,
which was introduced to transform the divergent UVDP-parameter integrals
into ILI-like one and the latter is the right object to be regularized in
the framework of LORE method. If this scale is left in the final result, we
must ask what is the physical meaning of this new scale and what the exact
value is. In fact, as argued in our previous paper\cite{Huang}, this scale
should be determined in terms of the intrinsic energy scales in the theory,
like the external momenta or particle masses. In the present case, the only
energy scale is the square of the external photon's momentum $p^{2}$, which,
in the massless electron case, needs to be space-like. Thus, if the $%
-q_{o}^{2}$ has some physical meaning, it must have some relation to $p^{2}$%
. The simplest possibility is that they are the same, $-q_{o}^{2}=-p^{2}$.
By applying the relation $-q_{o}^{2}=-p^{2}$, we can reach our final result
of photon vacuum polarization diagrams at two loop order:
\begin{eqnarray}
\mathcal{M}^{\text{final}} &\sim &\frac{8ie^{4}}{(16\pi ^{2})^{2}}(g^{\mu
\nu }p^{2}-p^{\mu }p^{\nu })(-\frac{1}{2})(\ln \frac{M_{c}^{2}}{-p^{2}}%
-\gamma _{\omega })  \notag  \label{final} \\
&=&\frac{i\alpha ^{2}}{4\pi ^{2}}(p^{\mu }p^{\nu }-g^{\mu \nu }p^{2})(\ln
\frac{M_{c}^{2}}{-p^{2}}-\gamma _{\omega }),
\end{eqnarray}%
where $\alpha =\frac{e^{2}}{4\pi ^{2}}$ is the fine structure constant.

In the light of our result, we have a few comments:

\begin{enumerate}
\item \noindent As we have already seen, the quadratic divergences are
canceled, and this is crucial to guarantee that photon does not obtain a
mass from quantum fluctuation and that the whole theory remains gauge
invariance.

\item The harmful divergences like $\log p^{2}\cdot \log M_{c}^{2}$ and $%
M_{c}^{2}\cdot \log p^{2}$ vanishes also, which is expected as these terms
are nonlocal and cannot be eliminated by any counterterms in the original
Lagrangian which are local. Especially, the quadratic harmful divergence
shows a new divergent structure, which can be expected \ from simple power
counting in each of these diagrams. The nontrivial thing is that for every
diagram this term will persist even when we include the corresponding
counterterm insertion diagrams which only cancel the logarithmically harmful
divergence. Only when we sum over all the relevant diagrams do the quadratic
harmful divergences cancel each other, rendering the final result local.%
\newline
Moreover, as emphasized in the previous subsections, this feature is also
expected from gauge invariance and the Ward identity in QED. Note that this
quadratically harmful divergence is only proportional to $g^{\mu \nu }$ and
can give the photon a mass term, both of which are forbidden by the gauge
invariance. We think this is a manifestation of the general phenomenon in
gauge theories like QCD that although each diagram contains the
gauge-violating quadratic divergences, but in the sum of all the relevant
diagrams they cancel. The only difference is that this phenomenon occurs at
two-loop order for QED while for QCD it happens at one-loop level already%
\cite{wu1,wu2}. It seems that it is the gauge invariance that rescues us
from the harmful divergence \textquotedblleft disaster\textquotedblright .

\item We can compare our calculation with the LORE method with those using
the dimensional regularization (DR). In the calculation with DR we cannot
see such cancellation of quadratic harmful divergences because in DR, the
final result comes from expansion around the actual space-time dimension 4
and in such a expansion only the logarithmic divergences can be preserved
while the quadratic or more higher divergences are simply omitted. It is the
resulting incomplete expression in DR that makes this new divergence
structure and its elimination invisible. However, since LORE enables us to
calculate the full results in the exactly four dimensions, we can easily
overcome this problem faced by DR, allowing the appearance of this
nontrivial divergence structure.

\item The final result Eq.(\ref{final}) is transverse, which implies that
our calculation respects gauge invariance and the corresponding Ward
Identity. Therefore, the final divergence can be canceled by a single
two-loop local counterterm for photon vacuum polarization.

\item Our result agrees with other authors' results \cite{Jost: 1950,
Itzykson:1980rh, bjor}.\newline
\end{enumerate}

\indent Therefore we come to our conclusion: LORE method, together with the
prescription of consistency conditions, can be safely applied to general
gauge theories, like QED, at least, up to two-loop order. 

\section{Conclusion}

In this paper, we mainly discuss the problem how to deal with the general
tensor-type high-loop integrals in the framework of LORE. We show that LORE,
together with its consistency conditions for 1-folded ILIs, are enough to
regularize the tensor-type integrals properly and can be consistently
applied to the general gauge theories. In order to show this conclusion, we
use the two-loop massless QED vacuum polarization as the simplest gauge
theory example to show the general procedure. The final result is sensible:
the two-loop correction is transverse and the harmful divergences and
quadratic divergences are naturally canceled, which are required by the
locality and the gauge invariance and its associated Ward identity of the
underlying QED. Consequently, our result agrees with other authors'
well-established one. Therefore, the computation of massless QED vacuum
polarization at two-loop order explicitly shows LORE is consistent with the
gauge invariance and give us confidence to apply it to other realistic
theories, like the standard model in particle physics.

What we want to emphasize is that from our present calculation, we have
shown a new divergence structure: either the self-energy insertion diagram $%
(a_{1})$ and $(a_{2})$ or the vertex correction insertion diagram $(b)$ may
contain gauge non-invariant quadratically harmful divergent term even when
combined with their own one-loop counterterm diagrams. However, only when we
sum over all diagrams in this order the harmful divergences cancel and the
final result recovers the gauge invariance and locality. We argue that this
feature is general for gauge theories, because the quadratic divergences
which violate the gauge invariance will give the gauge boson mass and should
be absent in the physical gauge invariant results. Furthermore, the reason
why LORE method, rather than other regularization methods like dimensional
regularization, can show this new divergence structure lies in the fact that
LORE enables us to calculate the complete expression for any Feynman
diagram, especially including the quadratic divergences, which is one of
main advantages of LORE.

\vspace{1 cm}

\centerline{{\bf Acknowledgement}}

\vspace{20 pt}

\noindent The authors would like to thank Jianwei Cui, Yibo Yang and Yong
Tang for useful discussions. This work was supported in part by the National
Science Foundation of China (NSFC) under Grant \#No. 10821504, 10975170 and
the Project of Knowledge Innovation Program (PKIP) of the Chinese Academy of
Science.

\newpage \appendix

\section{Useful Formula for the Regularization ILIs in LORE}

In the following, we will list some of the most useful regularized ILIs for
reference:\newline
\begin{eqnarray}
& & \int [d^4 k]_l \frac{1}{k^2 + M_l^2 + \mathcal{M}^2 } = \pi^2 \{ M_c^2 -
\mu^2 [\ \ln \frac{M_c^2}{\mu^2} - \gamma_w + 1 + y_2(\frac{\mu^2}{M_c^2}) \
] \} \\
& & \int [d^4 k]_l \frac{1}{ (k^2 + M_l^2 + \mathcal{M}^2 )^2} = \pi^2 \{ \
\ln \frac{M_c^2}{\mu^2} - \gamma_w + y_0(\frac{\mu^2}{M_c^2}) \ \} \\
& & \int [d^4 k]_l \frac{1}{ (k^2 + M_l^2 + \mathcal{M}^2)^3} = \frac{1}{%
2\mu^2} \pi^2 \{ \ 1- y_{-2}(\frac{\mu^2}{M_c^2}) \ \} \\
& & \int [d^4 k]_l \frac{1}{ (k^2 + M_l^2 + \mathcal{M}^2 )^{\alpha}} =
\pi^2 \frac{\Gamma(\alpha -2)}{\Gamma(\alpha) } \frac{1}{( \mu^2 )^{\alpha
-2} } \{ \ 1 - y_{-2(\alpha -2)} (\frac{\mu^2}{M_c^2}) \ \} \qquad \alpha > 3
\notag \\
\\
& & \int [d^4 k]_l \frac{k{\mu}k_{\nu}}{ (k^2 + M_l^2 + \mathcal{M}^2 )^2 }
= \frac{1}{2} \delta_{\mu\nu} \pi^2 \{ M_c^2 - \mu^2 [\ \ln \frac{M_c^2}{%
\mu^2} - \gamma_w + 1 + y_2(\frac{\mu^2}{M_c^2}) \ ] \} \\
& & \int [d^4 k]_l \frac{k{\mu}k_{\nu}}{ ( k^2 + M_l^2 + \mathcal{M}^2 )^3 }
= \frac{1}{4} \delta_{\mu\nu} \pi^2 \{ \ \ln \frac{M_c^2}{\mu^2} - \gamma_w
+ y_0(\frac{\mu^2}{M_c^2}) \ \} \\
& & \int [d^4 k]_l \frac{k{\mu}k_{\nu}}{ ( k^2 + M_l^2 + \mathcal{M}^2 )^4 }
= \frac{1}{4}\delta_{\mu\nu} \frac{1}{3\mu^2} \pi^2 \{ \ 1- y_{-2}(\frac{%
\mu^2}{M_c^2}) \ \} \\
& & \int [d^4 k]_l \frac{k{\mu}k_{\nu}}{ ( k^2 + M_l^2 + \mathcal{M}^2
)^{\alpha + 1} } = \frac{1}{2} \delta_{\mu\nu} \pi^2 \frac{\Gamma(\alpha -2)%
}{\Gamma(\alpha + 1) } \frac{1}{( \mu^2 )^{\alpha -2} } \{ \ 1 -
y_{-2(\alpha -2)} (\frac{\mu^2}{M_c^2}) \ \}  \notag \\
\end{eqnarray}
\begin{eqnarray}
& & \int [d^4 k]_l \frac{k{\mu}k_{\nu}k_{\rho}k_{\sigma}}{ (k^2 + M_l^2 +
\mathcal{M}^2 )^3 } = \frac{1}{8} \delta_{\{ \mu\nu\rho\sigma\} } \pi^2 \{
M_c^2 - \mu^2 [\ \ln \frac{M_c^2}{\mu^2} - \gamma_w + 1 + y_2(\frac{\mu^2}{%
M_c^2}) \ ] \}  \notag \\
\\
& & \int [d^4 k]_l \frac{k{\mu}k_{\nu}k_{\rho}k_{\sigma} }{ ( k^2 + M_l^2 +
\mathcal{M}^2 )^4 } = \frac{1}{24} \delta_{\{ \mu\nu\rho\sigma\} } \pi^2 \{
\ \ln \frac{M_c^2}{\mu^2} - \gamma_w + y_0(\frac{\mu^2}{M_c^2}) \ \} \\
& & \int [d^4 k]_l \frac{k{\mu}k_{\nu} k_{\rho}k_{\sigma} }{ (k^2 + M_l^2 +
\mathcal{M}^2 )^5 } = \frac{1}{24} \delta_{\{ \mu\nu\rho\sigma\} } \frac{1}{%
4\mu^2} \pi^2 \{ \ 1- y_{-2}(\frac{\mu^2}{M_c^2}) \ \} \\
& & \int [d^4 k]_l \frac{k{\mu}k_{\nu} k_{\rho}k_{\sigma} }{ ( k^2 + M_l^2 +
\mathcal{M}^2 )^{\alpha + 2} } = \frac{1}{4} \delta_{\{ \mu\nu\rho\sigma\} }
\pi^2 \frac{\Gamma(\alpha -2)}{\Gamma(\alpha + 2) } \frac{1}{( \mu^2
)^{\alpha -2} } \{ \ 1 - y_{-2(\alpha -2)} (\frac{\mu^2}{M_c^2}) \ \}  \notag
\\
\end{eqnarray}
with
\begin{eqnarray}
\int [d^4 k]_l \equiv \lim_{N\to \infty} \sum_{l=0}^{N} c_l^N \int d^4 k =
\lim_{N\to \infty} \sum_{l=0}^{N} (-1)^l \frac{N!}{(N-l)!\ l!} \int d^4 k
\end{eqnarray}
where
\begin{eqnarray}
& & M_l^2 = \mu_s^2 + l M_R^2 \ , \quad M_R^2 = M_c^2 h_w(N) \ln N \\
& & \mu^2 = \mu_s^2 + \mathcal{M}^2 \\
& & \delta_{\{ \mu\nu\rho\sigma\} } \equiv \delta_{\mu\nu}
\delta_{\rho\sigma} + \delta_{\mu\rho} \delta_{\nu\sigma} +
\delta_{\mu\sigma} \delta_{\rho\nu} \\
& & y_{-2(\alpha -2)} (x) = -\lim_N \sum_{l=1}^{N} c_l^N \left( \frac{x/(l
h_w(N) \ln N )}{ 1 + x/(l h_w(N) \ln N ) } \right)^{\alpha -2} \\
& & \gamma_w = \gamma_E = 0.5772 \cdots \ , \quad h_w(N\to\infty) = 1
\end{eqnarray}
The explicit form of $y_{-2(\alpha-2)}(x)$ is already given by Eq.(\ref%
{y-function}) 

\section{Details of Calculation of Self-Energy Insertion Diagram}

\label{AUVDP}

In this Appendix, we shall present the details of the calculation of the
UVDP parameter integrals for the self-energy correction insertion diagram $%
(a_1)$ shown in Fig.(\ref{qedA}).

With the simplification by setting $\mu_s^2\rightarrow0$ and $y_i(\frac{\mu^2%
}{M_c^2})\rightarrow0$ in Eq.(\ref{aa}), we have the following expression
for $\mathcal{M}^{(a_1)}_2$
\begin{eqnarray}
\mathcal{M}^{(a_1)}_2 &=& \frac{8ie^4\Gamma(4)}{(16\pi^2)^2}\int^\infty_0%
\frac{du}{1+u}\int^1_0dx \int^\infty_0\frac{dv_1v_1}{(1+v_1)^3}\frac{1}{6}%
g^{\mu\nu}  \notag \\
&& \{M_c^2+\frac{uxv_1}{(1+u)^2}(1-\frac{xv_1}{1+v_1})p^2[(\ln\frac{M_c^2}{%
-p^2}-\gamma_\omega+1)- \ln\Big(\frac{uxv_1}{(1+u)^2}(1-\frac{xv_1}{1+v_1})%
\Big)]\}  \notag \\
&=& \mathcal{M}^{(a_1)}_{22}+\mathcal{M}^{(a_1)}_{20} + \mathcal{M}%
^{(a_1)}_{2R}.
\end{eqnarray}
Here $\mathcal{M}^{(a_1)}_{22}$, $\mathcal{M}^{(a_1)}_{20}$ and $\mathcal{M}%
^{(a_1)}_{2R}$ represent quadratical divergent, logarithmical divergent, and
regular part in the $\mathcal{M}^{(a_1)}_2$.

\begin{eqnarray}  \label{aI_1}
\mathcal{M}^{(a_1)}_{22} &=& \frac{8ie^4\Gamma(4)}{(16\pi^2)^2}\frac{1}{6}%
g^{\mu\nu}M_c^2\int^\infty_0\frac{du}{1+u}\int^1_0dx \int^\infty_0\frac{%
dv_1v_1}{(1+v_1)^3}  \notag \\
&=& \frac{8ie^4}{(16\pi^2)^2}g^{\mu\nu}M_c^2\frac{1}{2}\int^\infty_0\frac{du%
}{1+u}  \notag \\
&=& \frac{4ie^4}{(16\pi^2)^2}g^{\mu\nu}M_c^2\int^\infty_0\frac{[dq_u^2]_l}{%
q_u^2-q_o^2}  \notag \\
&=& \frac{4ie^4}{(16\pi^2)^2}g^{\mu\nu}M_c^2[\ln\frac{M_c^2}{\mu^2_s-q_o^2}%
-\gamma_\omega+y_0(\frac{\mu_s^2-q_o^2}{M_c^2})]  \notag \\
&=& \frac{4ie^4}{(16\pi^2)^2}g^{\mu\nu}M_c^2[\ln\frac{M_c^2}{-q_o^2}%
-\gamma_\omega]
\end{eqnarray}
where in the third line we use the trick that $q_u^2\equiv-q_o^2 u$, which
can effectively transform a UVDP parameter integral into a 1-folded
irreducible loop integral(ILI) that is the fundamental object to be
regulated in the framework of LORE. In the following, we will frequently use
this trick without any explanation.
\begin{eqnarray}  \label{aI_2}
\mathcal{M}^{(a_1)}_{20} &=& \frac{8ie^4\Gamma(4)}{(16\pi^2)^2}\frac{1}{6}%
g^{\mu\nu}p^2(\ln\frac{M_c^2}{-p^2}-\gamma_\omega+1) \int^\infty_0\frac{du u%
}{(1+u)^3} \int^1_0dx \int^\infty_0\frac{dv_1 v_1}{(1+v_1)^3}xv_1(1-\frac{%
xv_1}{1+v_1})  \notag \\
&=& \frac{8ie^4}{(16\pi^2)^2}g^{\mu\nu}p^2(\ln\frac{M_c^2}{-p^2}%
-\gamma_\omega+1)\frac{1}{2} \int^\infty_0\frac{dv_1 v_1^2}{(1+v_1)^3}[\frac{%
1}{2}-\frac{v_1}{3(1+v_1)}]  \notag \\
&=& \frac{4ie^4}{(16\pi^2)^2}g^{\mu\nu}p^2(\ln\frac{M_c^2}{-p^2}%
-\gamma_\omega+1)\cdot (\frac{1}{6}\int^\infty_0\frac{dv_1}{1+v_1}-\frac{5}{%
36})  \notag \\
&=& \frac{4ie^4}{(16\pi^2)^2}g^{\mu\nu}p^2(\ln\frac{M_c^2}{-p^2}%
-\gamma_\omega+1)\cdot (\frac{1}{6}\int^\infty_0\frac{[dq_1^2]_l}{q_1^2-q_o^2%
}-\frac{5}{36})  \notag \\
&=& \frac{4ie^4}{(16\pi^2)^2}g^{\mu\nu}p^2(\ln\frac{M_c^2}{-p^2}%
-\gamma_\omega+1)\cdot [\frac{1}{6}(\ln\frac{M_c^2}{-q^2_o}-\gamma_\omega)-%
\frac{5}{36}]
\end{eqnarray}

The integral
\begin{eqnarray}
\mathcal{M}^{(a_1)}_{2R} &=& \frac{8ie^4\Gamma(4)}{(16\pi^2)^2}\frac{1}{6}%
g^{\mu\nu}(-p^2)\int^\infty_0\frac{du u}{(1+u)^3} \int^1_0dx\int^\infty_0%
\frac{dv_1 v_1}{(1+v_1)^3}  \notag \\
&&xv_1(1-\frac{xv_1}{1+v_1})\ln\Big(\frac{uxv_1}{(1+u)^2}(1-\frac{xv_1}{1+v_1%
})\Big)
\end{eqnarray}
is difficult since the complicated form in the logarithm. However, if we
only focus on its divergence behavior, we can make a lot of simplification.
As discussed in Section 2 in our previous paper\cite{Huang}, the analogy
between circuit diagram and Fig. (\ref{qedA}) tells us that the divergence
appear only in the region $v_1\rightarrow\infty$. Therefore, we can set $v_1$
in the region $v_1>V$, which $V$ is a very large number $V\gg 1$. In this
asymptotic region, we have $\frac{v_1}{1+v_1}\rightarrow1$ and the above
expression simplifies to:
\begin{eqnarray}
\mathcal{M}^{(a_1)}_{2R}&\sim& \frac{8ie^4\Gamma(4)}{(16\pi^2)^2}\frac{1}{6}%
g^{\mu\nu}(-p^2)\int^\infty_0\frac{du u}{(1+u)^3} \int^1_0dx x(1-x)
\int^\infty_V\frac{dv_1}{1+v_1}  \notag \\
&& [\ln\frac{u}{(1+u)^2}+\ln\Big(x(1-x)\Big)+\ln(1+v_1)]  \notag \\
&\equiv& \mathcal{M}^{(a_1)}_{2Ru}+\mathcal{M}^{(a_1)}_{2Rx}+\mathcal{M}%
^{(a_1)}_{2Rv_1},
\end{eqnarray}
where $\mathcal{M}^{(a_1)}_{2Ru}$, $\mathcal{M}^{(a_1)}_{2Rx}$ and $\mathcal{%
M}^{(a_1)}_{2Rv_1}$ represent the three parts in the second line,
respectively, the results of which are given as follows:
\begin{eqnarray}  \label{aI_3(1)}
\mathcal{M}^{(a_1)}_{2Ru} &=& \frac{8ie^4}{(16\pi^2)^2}(-g^{\mu\nu}p^2)\int^%
\infty_0\frac{du u}{(1+u)^3} \ln\frac{u}{(1+u)^2}\int^1_0dx
x(1-x)\int^\infty_V\frac{dv_1}{1+v_1}  \notag \\
&=& \frac{8ie^4}{(16\pi^2)^2}(-g^{\mu\nu}p^2)(-1)\frac{1}{6} \int^\infty_V%
\frac{dv_1}{1+v_1}  \notag \\
&=& \frac{4ie^4}{(16\pi^2)^2}\frac{1}{3}(g^{\mu\nu}p^2)(\ln\frac{M_c^2}{%
-q_o^2(1+V)}-\gamma_\omega),
\end{eqnarray}
\begin{eqnarray}  \label{aI_3(2)}
\mathcal{M}^{(a_1)}_{2Rx} &=& \frac{8ie^4}{(16\pi^2)^2}(-g^{\mu\nu}p^2)\int^%
\infty_0\frac{du u}{(1+u)^3} \int^1_0dx x(1-x)\ln[x(1-x)]\int^\infty_V\frac{%
dv_1}{1+v_1}  \notag \\
&=& \frac{8ie^4}{(16\pi^2)^2}(-g^{\mu\nu}p^2){\frac{1}{2}}(-{\frac{5}{18}}%
)\int^\infty_V\frac{dv_1}{1+v_1}  \notag \\
&=& \frac{4ie^4}{(16\pi^2)^2}\frac{5}{18}g^{\mu\nu}p^2(\ln\frac{M_c^2}{%
-q_o^2(1+V)}-\gamma_\omega),
\end{eqnarray}
\begin{eqnarray}  \label{aI_3(31)}
\mathcal{M}^{(a_1)}_{2Rv_1} &=& \frac{8ie^4}{(16\pi^2)^2}(-g^{\mu\nu}p^2)%
\int^\infty_0\frac{du u}{(1+u)^3} \int^1_0dx [x(1-x)]\int^\infty_V\frac{dv_1%
}{1+v_1}\ln(1+v_1)  \notag \\
&=& \frac{4ie^4}{(16\pi^2)^2}\frac{1}{6}(-g^{\mu\nu}p^2) \int^\infty_V\frac{%
dv_1}{1+v_1}\ln(1+v_1)  \notag \\
&=& \frac{4ie^4}{(16\pi^2)^2}\frac{1}{6}(-g^{\mu\nu}p^2) [\int^\infty_0\frac{%
dv_1}{1+v_1}\ln(1+v_1)-\int^\infty_V\frac{dv_1}{1+v_1}\ln(1+v_1)]
\end{eqnarray}
Since we are only interested in the divergence behavior, the last finite
term can be neglected due to the complexity of its calculation. Below we
only give the regulated result for the first term with LORE method and leave
the details of derivation to Appendix D.
\begin{eqnarray}  \label{aI_3(3)}
\mathcal{M}^{(a_1)}_{2Rv_1} &\sim& \frac{4ie^4}{(16\pi^2)^2}\frac{1}{6}%
(-g^{\mu\nu}p^2) \int^\infty_0\frac{dv_1}{1+v_1}\ln(1+v_1)  \notag \\
&=& \frac{4ie^4}{(16\pi^2)^2}\frac{1}{12}(-g^{\mu\nu}p^2) [(\ln\frac{M_c^2}{%
-q_o^2}-\gamma_\omega)^2-\alpha_\omega].
\end{eqnarray}

Putting Eq. (\ref{aI_3(1)}), (\ref{aI_3(2)}) and (\ref{aI_3(3)}) together,
we arrive at the result for $\mathcal{M}^{(a_1)}_{2R}$
\begin{eqnarray}  \label{aI_3}
\mathcal{M}^{(a_1)}_{2R} \sim \frac{4ie^4}{(16\pi^2)^2}g^{\mu\nu}p^2[\frac{11%
}{18}(\ln\frac{M_c^2}{-q_o^2(1+V)}-\gamma_\omega)- \frac{1}{12}(\ln\frac{%
M_c^2}{-q_o^2}-\gamma_\omega)^2+\frac{1}{12}\alpha_\omega].
\end{eqnarray}

With the experience of the calculation of $\mathcal{M}^{(a_1)}_{2}$, the
computation of logarithmic part $\mathcal{M}^{(a_1)}_{0}$ is
straightforward:
\begin{eqnarray}
\mathcal{M}^{(a_1)}_{0} &=& \frac{8ie^4\Gamma(4)}{(16\pi^2)^2}(-\frac{1}{6}%
)(2p^\mu p^\nu-g^{\mu\nu}p^2)\int^\infty_0\frac{du u}{(1+u)^3} \int^1_0dx
\int^\infty_0\frac{dv_1 v_1}{(1+v_1)^3} xv_1(1-\frac{xv_1}{1+v_1})  \notag \\
&& [(\ln\frac{M_c^2}{-p^2}-\gamma_\omega) -\ln\frac{uxv_1}{(1+u)^2}(1-\frac{%
xv_1}{1+v_1})]  \notag \\
&\equiv& \mathcal{M}^{(a_1)}_{00}+\mathcal{M}^{(a_1)}_{0R}.
\end{eqnarray}
>From the definition of $\mathcal{M}^{(a_1)}_{00(R)}$, it is easy to see
that they are essentially the same as counterparts in the quadratic part $%
\mathcal{M}^{(a_1)}_{20(R)}$. Thus, we can write down the final expression
for $\mathcal{M}^{(a_1)}_{00(R)}$ immediately following $\mathcal{M}%
^{(a_1)}_{20(R)}$.
\begin{equation}  \label{aII_1}
\mathcal{M}^{(a_1)}_{00} = \frac{4ie^4}{(16\pi^2)^2}(g^{\mu\nu}p^2-2p^\mu
p^\nu)(\ln\frac{M_c^2}{-p^2}-\gamma_\omega)\cdot [\frac{1}{6}(\ln\frac{M_c^2%
}{-q^2_o}-\gamma_\omega)-\frac{5}{36}]
\end{equation}
\begin{equation}  \label{aII_2}
\mathcal{M}^{(a_1)}_{0R} \sim \frac{4ie^4}{(16\pi^2)^2}(g^{\mu\nu}p^2-2p^\mu
p^\nu) [\frac{11}{18}(\ln\frac{M_c^2}{-q_o^2(1+V)}-\gamma_\omega)- \frac{1}{%
12}(\ln\frac{M_c^2}{-q_o^2}-\gamma_\omega)^2+\frac{1}{12}\alpha_\omega]
\end{equation}
By summing Eqs. (\ref{aI_1}), (\ref{aI_2}), (\ref{aI_3}), (\ref{aII_1}), (%
\ref{aII_2}), we finally obtain our result for diagram $(a_1)$
\begin{eqnarray}  \label{a_result}
\mathcal{M}^{(a_1)} &\sim& \frac{4ie^4}{(16\pi^2)^2}\{g^{\mu\nu}M_c^2(\ln%
\frac{M_c^2}{-q_o^2}-\gamma_\omega) +\frac{1}{6}g^{\mu\nu}p^2(\ln\frac{M_c^2%
}{-q_o^2}-\gamma_\omega)  \notag \\
&& +(g^{\mu\nu}p^2-p^\mu p^\nu)\cdot [\frac{1}{3}(\ln\frac{M_c^2}{-p^2}%
-\gamma_\omega)(\ln\frac{M_c^2}{-q_o^2}-\gamma_\omega) -\frac{1}{6}(\ln\frac{%
M_c^2}{-q_o^2}-\gamma_\omega)^2 +\frac{1}{6}\alpha_\omega  \notag \\
&&-\frac{5}{18}(\ln\frac{M_c^2}{-p^2}-\gamma_\omega) +\frac{11}{9}(\ln\frac{%
M_c^2}{-q_o^2}-\gamma_\omega)] \}
\end{eqnarray}

\section{Details of Calculation of Vertex Correction Insertion Diagram}

\label{BUVDP}

This appendix gives the details of the calculation of integration of UVDP
parameters for the vertex correction insertion diagram $(b)$.

Note that the integration of $k_2$ naturally separate $\mathcal{M}^{(b)}$
into four pieces with different powers of factor $\frac{1}{u+\frac{1}{%
(1+v_1)(1+v_2)}}$, each of which we denote as $\mathcal{M}^{(b)}_{i}$, $%
i=0,1,2,3$. In each piece, there are different degree of divergences,
quadratically or logarithmically divergent and regular. In the following, we
will use a second subscript $2,0,R$ to represent these parts. Let us compute
these parts one by one.

Firstly, $\mathcal{M}^{(b)}_{0}$ contains only the single quadratically
divergent term $\mathcal{M}^{(b)}_{02}$, whose result can be given directly
as:
\begin{eqnarray}  \label{b2}
\mathcal{M}^{(b)}_{02} &=& \frac{8ie^4\Gamma(4)}{16\pi^2}\int^\infty_0%
\prod^4_{i=1}\frac{dv_i}{(1+v_i)^2} \delta(1-\sum^4_{i=1}\frac{1}{1+v_i})
g^{\mu\nu}\frac{5}{6}I^{R}_2(0)  \notag \\
&=& \frac{8ie^4\Gamma(4)}{(16\pi^2)^2}\frac{5}{36}g^{\mu\nu}M_c^2,
\end{eqnarray}
where, as noted before, we have set $\mu_s^2\to0$ in the end since it can be
regarded as just the IR regularization in the LORE and does not affect the
UV structure of Feynman integrals. In the following, we will always apply
this method to simplify our results.

Also, $\mathcal{M}^{(b)}_{1}$ merely includes one term $\mathcal{M}%
^{(b)}_{12}$, but the calculation is a little involved since it can be
divided further to several pieces $\mathcal{M}^{(b)}_{122}$, $\mathcal{M}%
^{(b)}_{120}$ and $\mathcal{M}^{(b)}_{12R}$ according to the degree of
divergences each of them contain:
\begin{eqnarray}  \label{b1 exp}
\mathcal{M}^{(b)}_{12} &=& -\frac{8ie^4\Gamma(4)}{16\pi^2}%
\int^1_0\prod^2_{i=1}dx_i \int^\infty_0\prod^{2}_{i=1}\frac{dv_i}{(1+v_i)^3}%
\delta(1-\sum^2_{i=1}\frac{1}{1+v_i})\int^\infty_0 \frac{du}{u+\frac{1}{%
(1+v_1)(1+v_2)}}I^R_2  \notag \\
&=& -\frac{8ie^4\Gamma(4)}{(16\pi^2)^2}\int^1_0\prod^2_{i=1}dx_i\int^%
\infty_0\prod^{2}_{i=1}\frac{dv_i}{(1+v_i)^3} \delta(1-\sum^2_{i=1}\frac{1}{%
1+v_i})\int^\infty_0 \frac{du}{u+\frac{1}{(1+v_1)(1+v_2)}}  \notag \\
&& \{M_c^2-\mu_u^2[\ln\frac{M_c^2}{\mu_u^2}-\gamma_\omega+1]\}  \notag \\
&=& \mathcal{M}^{(b)}_{122}+\mathcal{M}^{(b)}_{120}+\mathcal{M}^{(b)}_{12R}.
\end{eqnarray}
Each term above can be calculated straightforwardly:
\begin{eqnarray}
\mathcal{M}^{(b)}_{122} &=& -\frac{8ie^4\Gamma(4)}{(16\pi^2)^2}%
\int^1_0\prod^2_{i=1}dx_i \int^\infty_0\prod^{2}_{i=1}\frac{dv_i}{(1+v_i)^3}%
\delta(1-\sum^2_{i=1}\frac{1}{1+v_i}) \int^\infty_0 \frac{du}{u+\frac{1}{%
(1+v_1)(1+v_2)}}M_c^2  \notag \\
&=& -\frac{8ie^4\Gamma(4)}{(16\pi^2)^2} M_c^2
\int^1_0\prod^2_{i=1}dx_i\int^\infty_0\prod^{2}_{i=1}\frac{dv_i}{(1+v_i)^3}%
\delta(1-\sum^2_{i=1}\frac{1}{1+v_i})\int^\infty_0 \frac{d q_u^2}{q_u^2-%
\frac{q_o^2}{(1+v_1)(1+v_2)}}  \notag \\
&=& -\frac{8ie^4\Gamma(4)}{(16\pi^2)^2}M_c^2\int^1_0\prod^2_{i=1}dx_i
\int^\infty_0\prod^{2}_{i=1} \frac{dv_i}{(1+v_i)^3} \delta(1-\sum^2_{i=1}%
\frac{1}{1+v_i})  \notag \\
&&[\ln\frac{M_c^2}{-q_o^2 \frac{1}{(1+v_1)(1+v_2)}} -\gamma_\omega]  \notag
\\
&=& -\frac{8ie^4\Gamma(4)}{(16\pi^2)^2}M_c^2[\frac{1}{6}(\ln\frac{M_c^2}{%
-q_o^2}-\gamma_\omega) +\frac{5}{18}],
\end{eqnarray}
\begin{eqnarray}
\mathcal{M}^{(b)}_{{120}} &=& -\frac{8ie^4\Gamma(4)}{(16\pi^2)^2} (\ln\frac{%
M_c^2}{-p^2}-\gamma_\omega)
\int^1_0\prod^2_{i=1}dx_i\int^\infty_0\prod^{2}_{i=1}\frac{dv_i}{(1+v_i)^3}
\delta(1-\sum^2_{i=1}\frac{1}{1+v_i}) \int^\infty_0du  \notag \\
&& \frac{1}{u+\frac{1}{(1+v_1)(1+v_2)}} \{\frac{1}{u+\frac{1}{(1+v_1)(1+v_2)}%
}[\frac{(x_2-x_1)^2}{(1+v_1)(1+v_2)} +\frac{x_1(1-x_1)}{1+v_1}+\frac{%
x_2(1-x_2)}{1+v_2}]  \notag \\
&&-\frac{1}{[u+\frac{1}{(1+v_1)(1+v_2)}]^2} \frac{(x_2-x_1)^2}{%
(1+v_1)^2(1+v_2)^2}\}p^2  \notag \\
&=& -\frac{8ie^4\Gamma(4)}{(16\pi^2)^2}p^2\frac{13}{72}(\ln\frac{M_c^2}{-p^2}%
-\gamma_\omega).
\end{eqnarray}
Since it is easy to prove that $\mathcal{M}^{(b)}_{12R}$ is finite and thus
irrelevant to our present discussion of UV divergence structure, we omit its
calculation here and simply write down the result for $\mathcal{M}%
^{(b)}_{12} $ as:
\begin{eqnarray}  \label{b1}
\mathcal{M}^{(b)}_{12} &=& \mathcal{M}^{(b)}_{122}+\mathcal{M}^{(b)}_{120}+%
\mathcal{M}^{(b)}_{12R}  \notag \\
&\sim& -\frac{8ie^4\Gamma(4)}{(16\pi^2)^2}[\frac{1}{6}M_c^2(\ln\frac{M_c^2}{%
-q_o^2}-\gamma_\omega)+\frac{5}{18}M_c^2+ \frac{13}{72}p^2(\ln\frac{M_c^2}{%
-p^2}-\gamma_\omega)],
\end{eqnarray}
where $\sim$ means that they are equal up to the divergence part.

By observation, $\mathcal{M}^{(b)}_{2}$ contains quadratically and
logarithmically divergent parts, $\mathcal{M}^{(b)}_{22}$ and $\mathcal{M}%
^{(b)}_{20}$. The computation of $\mathcal{M}^{(b)}_{22}$ is much like that
of $\mathcal{M}^{(b)}_{12}$ and the result is shown below:
\begin{eqnarray}
\mathcal{M}^{(b)}_{22} &=& -\frac{8ie^4\Gamma(4)}{(16\pi^2)^2}g^{\mu\nu}
\int^1_0\prod^2_{i=1}dx_i\int^\infty_0\prod^{2}_{i=1}\frac{dv_i}{(1+v_i)^3}
\delta(1-\sum^2_{i=1}\frac{1}{1+v_i}) \int^\infty_0du  \notag \\
&& \frac{-\frac{1}{(1+v_1)(1+v_2)}}{[u+\frac{1}{(1+v_1)(1+v_2)}]^2}
\{M_c^2-\mu_u^2[\ln\frac{M_c^2}{\mu_u^2}-\gamma_\omega+1]\}  \notag \\
&=& \mathcal{M}^{(b)}_{222}+\mathcal{M}^{(b)}_{220}+\mathcal{M}^{(b)}_{22R}
\end{eqnarray}
\begin{eqnarray}
\mathcal{M}^{(b)}_{222} &=& -\frac{8ie^4\Gamma(4)}{(16\pi^2)^2}g^{\mu\nu}
\int^1_0\prod^2_{i=1}dx_i\int^\infty_0\prod^{2}_{i=1}\frac{dv_i}{(1+v_i)^3}
\delta(1-\sum^2_{i=1}\frac{1}{1+v_i}) \int^\infty_0du  \notag \\
&& \frac{-\frac{1}{(1+v_1)(1+v_2)}}{[u+\frac{1}{(1+v_1)(1+v_2)}]^2}M_c^2
\notag \\
&=& \frac{8ie^4\Gamma(4)}{(16\pi^2)^2}g^{\mu\nu}M_c^2
\end{eqnarray}
\begin{eqnarray}
\mathcal{M}^{(b)}_{220} &=& -\frac{8ie^4\Gamma(4)}{(16\pi^2)^2}%
g^{\mu\nu}\int^1_0\prod^2_{i=1}dx_i \int^\infty_0\prod^{2}_{i=1}\frac{dv_i}{%
(1+v_i)^3} \delta(1-\sum^2_{i=1}\frac{1}{1+v_i}) \int^\infty_0du  \notag \\
&& \frac{\frac{1}{(1+v_1)(1+v_2)}}{[u+\frac{1}{(1+v_1)(1+v_2)}]^2} \{\frac{1%
}{u+\frac{1}{(1+v_1)(1+v_2)}}[\frac{(x_2-x_1)^2}{(1+v_1)(1+v_2)} +\frac{%
x_1(1-x_1)}{1+v_1}+\frac{x_2(1-x_2)}{1+v_2}]  \notag \\
&&-\frac{1}{[u+\frac{1}{(1+v_1)(1+v_2)}]^2} \frac{(x_2-x_1)^2}{%
(1+v_1)^2(1+v_2)^2}\} (-p^2)(\ln\frac{M_c^2}{-p^2}-\gamma_\omega)  \notag \\
&=& \frac{8ie^4\Gamma(4)}{(16\pi^2)^2}g^{\mu\nu}p^2\frac{19}{36\cdot6}(\ln%
\frac{M_c^2}{-p^2}-\gamma_\omega)
\end{eqnarray}
We can prove that the term $\mathcal{M}^{(b)}_{22R}$ is finite. So we have
the divergence behavior of the integration $\mathcal{M}^{(b)}_{22}$ is:
\begin{equation}  \label{b3}
\mathcal{M}^{(b)}_{22}\sim \frac{8ie^4\Gamma(4)}{(16\pi^2)^2}g^{\mu\nu} [%
\frac{1}{6}M_c^2+\frac{19}{36\cdot6}p^2(\ln\frac{M_c^2}{-p^2}-\gamma_\omega)]
\end{equation}

$\mathcal{M}^{(b)}_{20}$ contains many different Lorentz structures, so it
is useful to further refine it into three parts according to their Lorentz
structure: $\mathcal{M}^{(b)}_{201}$ represents terms proportional to $p^\mu
p^\nu$, $\mathcal{M}^{(b)}_{202}$ proportional to $g^{\mu\nu} p^2$ and $%
\mathcal{M}^{b}_{203}$ proportional to $g^{\mu\nu}\mu_u^2$.
\begin{eqnarray}
\mathcal{M}^{(b)}_{201} &=& -\frac{8ie^4\Gamma(4)}{(16\pi^2)^2}p^\mu p^\nu
\int^1_0\prod^2_{i=1}dx_i\int^\infty_0\prod^{2}_{i=1}\frac{dv_i}{(1+v_i)^3}
\delta(1-\sum^2_{i=1}\frac{1}{1+v_i}) \int^\infty_0du  \notag \\
&& \frac{1}{3}\frac{1}{[u+\frac{1}{(1+v_1)(1+v_2)}]^2} [\frac{(x_2-x_1)^2}{%
(1+v_1)^2(1+v_2)^2[u+\frac{1}{(1+v_1)(1+v_2)}]^2}-1] (\ln\frac{M_c^2}{\mu_u^2%
}-\gamma_\omega)  \notag \\
&=& \mathcal{M}^{(b)}_{2010}+\mathcal{M}^{(b)}_{201R},
\end{eqnarray}
where the last subscript of $\mathcal{M}^{(b)}_{2010(R)}$ represents the
divergence degree of the part.

\begin{eqnarray}
\mathcal{M}^{(b)}_{2010} &=& -\frac{8ie^4\Gamma(4)}{(16\pi^2)^2} p^\mu p^\nu
\int^1_0\prod^2_{i=1}dx_i\int^\infty_0\prod^{2}_{i=1}\frac{dv_i}{(1+v_i)^3}
\delta(1-\sum^2_{i=1}\frac{1}{1+v_i}) \int^\infty_0du  \notag \\
&& \frac{1}{3}\frac{1}{[u+\frac{1}{(1+v_1)(1+v_2)}]^2} [\frac{(x_2-x_1)^2}{%
(1+v_1)^2(1+v_2)^2[u+\frac{1}{(1+v_1)(1+v_2)}]^2}-1] (\ln\frac{M_c^2}{-p^2}%
-\gamma_\omega)  \notag \\
&=& \frac{8ie^4\Gamma(4)}{(16\pi^2)^2} \frac{17}{54} p^\mu p^\nu (\ln\frac{%
M_c^2}{-p^2}-\gamma_\omega).
\end{eqnarray}
As before, we can prove that the part $\mathcal{M}^{(b)}_{201R}$ is finite,
so the divergent part for $\mathcal{M}^{(b)}_{201}$ is:
\begin{eqnarray}  \label{b4}
\mathcal{M}^{(b)}_{201} \sim \frac{8ie^4\Gamma(4)}{(16\pi^2)^2} \frac{17}{54}
p^\mu p^\nu (\ln\frac{M_c^2}{-p^2}-\gamma_\omega).
\end{eqnarray}

By the very similar way, we can obtain the result of $\mathcal{M}%
^{(b)}_{202} $.
\begin{eqnarray}
\mathcal{M}^{(b)}_{202} &=& \frac{8ie^4\Gamma(4)}{(16\pi^2)^2}\cdot\frac{1}{6%
}g^{\mu\nu}p^2 \int^1_0\prod^2_{i=1}dx_i\int^\infty_0\prod^{2}_{i=1}\frac{%
dv_i}{(1+v_i)^3} \delta(1-\sum^2_{i=1}\frac{1}{1+v_i}) \int^\infty_0du
\notag \\
&& \frac{1}{[u+\frac{1}{(1+v_1)(1+v_2)}]^2} \{\frac{[1-\frac{6}{%
(1+v_1)(1+v_2)}]\frac{(x_2-x_1)^2}{(1+v_1)^2(1+v_2)^2}}{[u+\frac{1}{%
(1+v_1)(1+v_2)}]^2}  \notag \\
&&+6(\frac{x_1}{1+v_1}+\frac{x_2}{1+v_2}-\frac{1}{2}) (\frac{1}{1+v_1}-\frac{%
1}{1+v_2}) \frac{x_2-x_1}{(1+v_1)(1+v_2)} \frac{1}{u+\frac{1}{(1+v_1)(1+v_2)}%
}  \notag \\
&&-6(\frac{x_1}{1+v_1}+\frac{x_2}{1+v_2})(1-\frac{x_1}{1+v_1}-\frac{x_2}{%
1+v_2})\}(\ln\frac{M_c^2}{\mu_u^2}-\gamma_\omega)  \notag \\
&=& \mathcal{M}^{(b)}_{2020}+\mathcal{M}^{(b)}_{202R}
\end{eqnarray}
By integrating the expression before $\frac{1}{16\pi^2}(\ln\frac{M_c^2}{-p^2}%
-\gamma_\omega)$, we obtain $\mathcal{M}^{(b)}_{2020}$
\begin{eqnarray}
\mathcal{M}^{(b)}_{2020} &=& -\frac{8ie^4\Gamma(4)}{(16\pi^2)^2}\cdot\frac{5%
}{24}g^{\mu\nu}p^2(\ln\frac{M_c^2}{-p^2}-\gamma_\omega).
\end{eqnarray}
The part of $\mathcal{M}^{(b)}_{202R}$ can also be proved to be finite, so
the divergence part of $\mathcal{M}^{(b)}_{202}$ is:
\begin{eqnarray}  \label{b6}
\mathcal{M}^{(b)}_{202} &\sim& -\frac{8ie^4\Gamma(4)}{(16\pi^2)^2}\cdot\frac{%
5}{24}g^{\mu\nu}p^2(\ln\frac{M_c^2}{-p^2}-\gamma_\omega).
\end{eqnarray}

Much like $\mathcal{M}^{(b)}_{201}$ and $\mathcal{M}^{(b)}_{202}$ in
structure, it is expected that the $\mathcal{M}^{(b)}_{203}$ can be
calculated similarly. However, a very important new feature will appear: the
overlapping divergence structure hidden in our current LORE procedure.
\begin{eqnarray}
\mathcal{M}^{(b)}_{203} &=& \frac{8ie^4\Gamma(4)}{(16\pi^2)^2}g^{\mu\nu}p^2
\int^1_0\prod^2_{i=1}dx_i\int^\infty_0\prod^{2}_{i=1}\frac{dv_i}{(1+v_i)^3}
\delta(1-\sum^2_{i=1}\frac{1}{1+v_i}) \int^\infty_0du  \notag \\
&&\frac{1}{[u+\frac{1}{(1+v_1)(1+v_2)}]^2}[-\frac{1}{6}+\frac{1}{%
(1+v_1)(1+v_2)}] \cdot  \notag \\
&&\{\frac{1}{u+\frac{1}{(1+v_1)(1+v_2)}}[\frac{(x_2-x_1)^2}{(1+v_1)(1+v_2)} +%
\frac{x_1(1-x_1)}{1+v_1}+\frac{x_2(1-x_2)}{1+v_2}]  \notag \\
&&-\frac{1}{[u+\frac{1}{(1+v_1)(1+v_2)}]^2} \frac{(x_2-x_1)^2}{%
(1+v_1)^2(1+v_2)^2}\}\cdot  \notag \\
&&\Big(\lbrack \ln\frac{M_c^2}{-p^2}-\gamma_\omega]-  \notag \\
&&\ln\{\frac{1}{u+\frac{1}{(1+v_1)(1+v_2)}}[\frac{(x_2-x_1)^2}{(1+v_1)(1+v_2)%
} +\frac{x_1(1-x_1)}{1+v_1}+\frac{x_2(1-x_2)}{1+v_2}]  \notag \\
&&-\frac{1}{[u+\frac{1}{(1+v_1)(1+v_2)}]^2} \frac{(x_2-x_1)^2}{%
(1+v_1)^2(1+v_2)^2}\}\Big)  \notag \\
&=& \mathcal{M}^{(b)}_{2030}+\mathcal{M}^{(b)}_{203R},
\end{eqnarray}
where, $\mathcal{M}^{(b)}_{2030}$ represents the part proportional to $(\ln%
\frac{M_c^2}{-p^2}-\gamma_\omega)$ while $\mathcal{M}^{(b)}_{203R}$ for the
rest regular terms. The integration of $\mathcal{M}^{(b)}_{2030}$ is
straightforward.
\begin{eqnarray}  \label{b5_0}
\mathcal{M}^{(b)}_{2030} &=& \frac{8ie^4\Gamma(4)}{(16\pi^2)^2}g^{\mu\nu}p^2
\int^1_0\prod^2_{i=1}dx_i\int^\infty_0\prod^{2}_{i=1}\frac{dv_i}{(1+v_i)^3}
\delta(1-\sum^2_{i=1}\frac{1}{1+v_i}) \int^\infty_0du  \notag \\
&&\frac{1}{[u+\frac{1}{(1+v_1)(1+v_2)}]^2}g^{\mu\nu}[-\frac{1}{6}+\frac{1}{%
(1+v_1)(1+v_2)}] (\ln\frac{M_c^2}{-p^2}-\gamma_\omega)  \notag \\
&&\{\frac{1}{u+\frac{1}{(1+v_1)(1+v_2)}}[\frac{(x_2-x_1)^2}{(1+v_1)(1+v_2)} +%
\frac{x_1(1-x_1)}{1+v_1}+\frac{x_2(1-x_2)}{1+v_2}]  \notag \\
&&-\frac{1}{[u+\frac{1}{(1+v_1)(1+v_2)}]^2} \frac{(x_2-x_1)^2}{%
(1+v_1)^2(1+v_2)^2}\}  \notag \\
&=& \frac{8ie^4\Gamma(4)}{(16\pi^2)^2}g^{\mu\nu}p^2 (\ln\frac{M_c^2}{-p^2}%
-\gamma_\omega)[\frac{1}{12}- \frac{1}{72}(\int^\infty_0\frac{dv_1}{1+v_1}%
+\int^\infty_0\frac{dv_2}{1+v_2})]  \notag \\
&=& \frac{8ie^4\Gamma(4)}{(16\pi^2)^2}g^{\mu\nu}p^2 (\ln\frac{M_c^2}{-p^2}%
-\gamma_\omega)[\frac{1}{12} -\frac{1}{36}(\ln\frac{M_c^2}{-q_o^2}%
-\gamma_\omega)].
\end{eqnarray}
Note that the above parameter integration before $(\ln\frac{M_c^2}{-p^2}%
-\gamma_\omega)$ is also divergent, which is the signal of overlapping
divergences. It is clearer when we use the analogy between the Feynman
diagrams and the electric circuits, which tells us that the integration of $%
v_1$ and $v_2$ should reproduce the divergences coming from the subdiagrams
of left and right vertex corrections in Fig.(\ref{qedB}) since $\frac{1}{%
1+v_1}$ ($\frac{1}{1+v_2}$) times the effective propagator $%
k_1^2+x_1(1-x_1)p^2$($[(k_1+k_2+(x_2-x_1)p)]^2+x_2(1-x_2)p^2$) for the
left(right) circle in Eq.(\ref{combination}). When $v_1$($v_2$) tends to
infinity, the left effective propagator $\frac{1}{1+v_1}%
[k_1^2+x_1(1-x_1)p^2] $ (the right counterpart $\frac{1}{1+v_2}%
\{[(k_1+k_2+(x_2-x_1)p)]^2+x_2(1-x_2)p^2\}$) approaches zero, which means
that the left(right) half circle collapse to a point and that sub-circuit is
short-cut in the electric circuit language. This singular behavior will
become manifest as the divergence in the final integration of UVDP parameter
$v_1$($v_2$), just as the ones shown in Eq.(\ref{b5_0}) above. Thus, the
result in Eq.(\ref{b5_0}) can be understood as the subdivergences coming
from left and right vertex correction times the overall one $(\ln\frac{M_c^2%
}{-p^2}-\gamma_\omega)$, which is the definition of overlapping divergence
\cite{Peskin:1995ev, Huang}.

This overlapping divergence take us some further difficulties that the
integration of $\mathcal{M}^{(b)}_{203R}$ turn out to be divergent, which
gives us further contributions to our UV-divergence structure.

\begin{eqnarray}
\mathcal{M}^{(b)}_{203R}&=& -\frac{8ie^4\Gamma(4)}{(16\pi^2)^2}g^{\mu\nu}p^2
\int^1_0\prod^2_{i=1}dx_i\int^\infty_0\prod^{2}_{i=1}\frac{dv_i}{(1+v_i)^3}
\delta(1-\sum^2_{i=1}\frac{1}{1+v_i}) \int^\infty_0du  \notag \\
&&\frac{1}{[u+\frac{1}{(1+v_1)(1+v_2)}]^2}[-\frac{1}{6}+\frac{1}{%
(1+v_1)(1+v_2)}]  \notag \\
&&\{\frac{1}{u+\frac{1}{(1+v_1)(1+v_2)}}[\frac{(x_2-x_1)^2}{(1+v_1)(1+v_2)} +%
\frac{x_1(1-x_1)}{1+v_1}+\frac{x_2(1-x_2)}{1+v_2}]  \notag \\
&&-\frac{1}{[u+\frac{1}{(1+v_1)(1+v_2)}]^2} \frac{(x_2-x_1)^2}{%
(1+v_1)^2(1+v_2)^2}\}  \notag \\
&& \ln\{\frac{1}{u+\frac{1}{(1+v_1)(1+v_2)}}[\frac{(x_2-x_1)^2}{%
(1+v_1)(1+v_2)} +\frac{x_1(1-x_1)}{1+v_1}+\frac{x_2(1-x_2)}{1+v_2}]  \notag
\\
&&-\frac{1}{[u+\frac{1}{(1+v_1)(1+v_2)}]^2} \frac{(x_2-x_1)^2}{%
(1+v_1)^2(1+v_2)^2}\}.
\end{eqnarray}
Because of the complication in the logarithmic function, it is difficult to
get a closed analytical expression for this kinds of integration. However,
if we only focus the divergence behavior of the integral, then we can use
the method introduced in the calculation of $\mathcal{M}^{(a_1)}_{2R}$ to
greatly simplify the integral and to obtain the asymptotical results. To
disentangle the overlapping divergences, we need first to know in what
parameter space region the divergence happen. From our experience when
working $\mathcal{M}^{(b)}_{2030}$, the divergences take place when $%
v_1\rightarrow \infty$ and $v_2\rightarrow \infty$.

Now we consider the region where $v_1\rightarrow\infty$ and $v_2\rightarrow0$%
. Fist we choose a very large number, say $V$, and set the integration
region is only confined in $v_1\gg V$. In such a region, $\frac{1}{1+v_1}$
are small quantities, so we can expand the expression according to $\frac{1}{%
1+v_1}$. The leading term of $\mathcal{M}^{(b)}_{203R}$ is,
\begin{eqnarray}  \label{b5_o-simp}
&&\mathcal{M}^{(b)}_{203Rv_1} \sim \frac{8ie^4\Gamma(4)}{(16\pi^2)^2}\cdot
\frac{1}{6}g^{\mu\nu}p^2 \int^1_0\prod^2_{i=1}dx_i \int^\infty_V \frac{dv_1}{%
(1+v_1)^3} \int^\infty_0 du \frac{x_2(1-x_2)}{(u+\frac{1}{1+v_1})^3} \ln%
\frac{x_2(1-x_2)}{u+\frac{1}{1+v_1}}  \notag \\
&=& \frac{8ie^4\Gamma(4)}{(16\pi^2)^2}\cdot \frac{1}{6}g^{\mu\nu}p^2
\{\int^1_0dx_2 x_2(1-x_2) \ln[x_2(1-x_2)]\int^\infty_V \frac{dv_1}{(1+v_1)^3}
\int^\infty_0 du \frac{1}{(u+\frac{1}{1+v_1})^3}  \notag \\
&&-\int^1_0dx_2 x_2(1-x_2) \int^\infty_V \frac{dv_1}{(1+v_1)^3}
\int^\infty_0 du \frac{1}{(u+\frac{1}{1+v_1})^3 }\ln [u+\frac{1}{1+v_1}]\}
\notag \\
&=& \frac{8ie^4\Gamma(4)}{(16\pi^2)^2}\cdot \frac{1}{6}g^{\mu\nu}p^2\{(-%
\frac{5}{18})\frac{1}{2} \int^\infty_V\frac{dv_1}{1+v_1} -\frac{1}{6}[\frac{1%
}{4}\int^\infty_V\frac{dv_1}{1+v_1}- \frac{1}{2}\int^\infty_V\frac{dv_1}{%
1+v_1}\ln(1+v_1)]\}  \notag \\
&\sim&-\frac{8ie^4\Gamma(4)}{(16\pi^2)^2}\cdot\frac{1}{72}g^{\mu\nu}p^2 [%
\frac{13}{6}(\ln\frac{M_c^2}{-q^2_o(1+V)}-\gamma_\omega) - \frac{1}{2}(\ln%
\frac{M_c^2}{-q^2_o}-\gamma_\omega)^2+\frac{1}{2}\alpha_\omega].
\end{eqnarray}
where in the last line we extended the lower bound of the integration range
to 0 as before for convenience.

\indent Since in the other asymptotic region $v_2\rightarrow\infty$, $%
v_1\rightarrow0$, we can obtain a similar expression except for the exchange
of $v_1 \leftrightarrow v_2$ and $x_2\leftrightarrow x_1$. Thus, we can
expect to get the same asymptotic result. Therefore, the divergence behavior
of $\mathcal{M}^{(b)}_{203R}$ is:
\begin{eqnarray}  \label{b5_o}
\mathcal{M}^{(b)}_{203R} \sim -\frac{8ie^4\Gamma(4)}{(16\pi^2)^2} \cdot\frac{%
1}{36}g^{\mu\nu}p^2 [\frac{13}{6}(\ln\frac{M_c^2}{-q^2_o(1+V)}%
-\gamma_\omega) - \frac{1}{2}(\ln\frac{M_c^2}{-q^2_o}-\gamma_\omega)^2+\frac{%
1}{2}\alpha_\omega].
\end{eqnarray}

$\mathcal{M}^{(b)}_{3}$ can be naturally divided into three parts $\mathcal{M%
}^{(b)}_{32}$, $\mathcal{M}^{(b)}_{30}$ and $\mathcal{M}^{(b)}_{3R}$
according to the divergence degree followed by integration of loop momentum $%
k_2$. $\mathcal{M}^{(b)}_{32}$ contains only one term, so the calculation is
straightforward:
\begin{eqnarray}
\mathcal{M}^{(b)}_{32} &=& -\frac{8ie^4\Gamma(4)}{(16\pi^2)^2}\cdot {\frac{1%
}{3}}g^{\mu\nu}\int^1_0\prod^2_{i=1}dx_i\int^\infty_0\prod^{2}_{i=1}\frac{%
dv_i}{(1+v_i)^3} \delta(1-\sum^2_{i=1}\frac{1}{1+v_i}) \int^\infty_0du
\notag \\
&&\frac{1}{[u+\frac{1}{(1+v_1)(1+v_2)}]^3}\frac{1}{(1+v_1)^2(1+v_2)^2}
[M_c^2-\mu_u^2(\ln\frac{M_c^2}{\mu_u^2}-\gamma_\omega+1)]  \notag \\
&=& \mathcal{M}^{(b)}_{322}+\mathcal{M}^{(b)}_{320}+\mathcal{M}^{(b)}_{32R}
\end{eqnarray}
\begin{eqnarray}
\mathcal{M}^{(b)}_{322} &=& -\frac{8ie^4\Gamma(4)}{(16\pi^2)^2}\cdot {\frac{1%
}{3}}g^{\mu\nu} M_c^2\int^1_0\prod^2_{i=1}dx_i\int^\infty_0\prod^{2}_{i=1}%
\frac{dv_i}{(1+v_i)^3} \delta(1-\sum^2_{i=1}\frac{1}{1+v_i}) \int^\infty_0du
\notag \\
&&\frac{1}{[u+\frac{1}{(1+v_1)(1+v_2)}]^3}\frac{1}{(1+v_1)^2(1+v_2)^2}
\notag \\
&=& -\frac{8ie^4\Gamma(4)}{(16\pi^2)^2}\frac{1}{36}g^{\mu\nu}M_c^2
\end{eqnarray}
\begin{eqnarray}
\mathcal{M}^{(b)}_{320} &=& -\frac{8ie^4\Gamma(4)}{(16\pi^2)^2} {\frac{1}{3}}%
g^{\mu\nu}p^2\int^1_0\prod^2_{i=1}dx_i\int^\infty_0\prod^{2}_{i=1}\frac{dv_i%
}{(1+v_i)^3} \delta(1-\sum^2_{i=1}\frac{1}{1+v_i}) \int^\infty_0du  \notag \\
&&\frac{1}{[u+\frac{1}{(1+v_1)(1+v_2)}]^3}\frac{1}{(1+v_1)^2(1+v_2)^2} (\ln%
\frac{M_c^2}{-p^2}-\gamma_\omega+1)  \notag \\
&& \{\frac{1}{u+\frac{1}{(1+v_1)(1+v_2)}}[\frac{(x_2-x_1)^2}{(1+v_1)(1+v_2)}
+\frac{x_1(1-x_1)}{1+v_1}+\frac{x_2(1-x_2)}{1+v_2}]  \notag \\
&&-\frac{1}{[u+\frac{1}{(1+v_1)(1+v_2)}]^2} \frac{(x_2-x_1)^2}{%
(1+v_1)^2(1+v_2)^2}\}  \notag \\
&=& -\frac{8ie^4\Gamma(4)}{(16\pi^2)^2}\frac{25}{36\cdot36}g^{\mu\nu}p^2(\ln%
\frac{M_c^2}{-p^2}-\gamma_\omega+1)
\end{eqnarray}
Since we can prove that $\mathcal{M}^{(b)}_{32R}$ is finite, the divergence
structure we concern now is given by the addition of $\mathcal{M}%
^{(b)}_{322} $ and $\mathcal{M}^{(b)}_{320}$:
\begin{eqnarray}  \label{b7}
\mathcal{M}^{(b)}_{32} \sim -\frac{8ie^4\Gamma(4)}{(16\pi^2)^2}\frac{1}{36}%
g^{\mu\nu}[M_c^2+ \frac{25}{36}p^2(\ln\frac{M_c^2}{-p^2}-\gamma_\omega+1)]
\end{eqnarray}
Like $\mathcal{M}^{(b)}_{20}$, $\mathcal{M}^{(b)}_{30}$ contains three parts
differentiating by their Lorentz structures: $\mathcal{M}^{(b)}_{301}$
represents part proportional to $p^{\mu} p^\nu$, $\mathcal{M}^{(b)}_{301}$
for $g^{\mu\nu}p^2$ and $\mathcal{M}^{(b)}_{303}$ for $g^{\mu\nu}\mu_u^2$.
Let us first calculate $\mathcal{M}^{(b)}_{301}$.
\begin{eqnarray}
\mathcal{M}^{(b)}_{301} &=& -\frac{8ie^4\Gamma(4)}{(16\pi^2)^2}\cdot\frac{1}{%
3}p^\mu p^\nu \int^1_0\prod^2_{i=1}dx_i\int^\infty_0\prod^{2}_{i=1}\frac{dv_i%
}{(1+v_i)^3} \delta(1-\sum^2_{i=1}\frac{1}{1+v_i}) \int^\infty_0du  \notag \\
&& [\frac{1}{(1+v_1)(1+v_2)} -(\frac{x_1}{1+v_1}+\frac{x_2}{1+v_2})(1-\frac{%
x_1}{1+v_1}-\frac{x_2}{1+v_2})]  \notag \\
&&\frac{1}{[u+\frac{1}{(1+v_1)(1+v_2)}]^3} (\ln\frac{M_c^2}{\mu_u^2}%
-\gamma_\omega)  \notag \\
&=& \mathcal{M}^{(b)}_{3010}+\mathcal{M}^{(b)}_{301R}.
\end{eqnarray}

Actual calculation of $\mathcal{M}^{(b)}_{3010}$ shows that the integral
before $(\ln\frac{M_c^2}{p^2}-\gamma_\omega)$ is logarithmically divergent.
So like the part $\mathcal{M}^{(b)}_{2030}$, $\mathcal{M}^{(b)}_{3010}$ also
involves overlapping divergences.

\begin{eqnarray}  \label{b10_0}
\mathcal{M}^{(b)}_{3010} &=& -\frac{8ie^4\Gamma(4)}{(16\pi^2)^2}\cdot \frac{1%
}{3}p^\mu p^\nu \int^1_0\prod^2_{i=1}dx_i\int^\infty_0\prod^{2}_{i=1}\frac{%
dv_i}{(1+v_i)^3} \delta(1-\sum^2_{i=1}\frac{1}{1+v_i}) \int^\infty_0du
\notag \\
&& [\frac{1}{(1+v_1)(1+v_2)} -(\frac{x_1}{1+v_1}+\frac{x_2}{1+v_2})(1-\frac{%
x_1}{1+v_1}-\frac{x_2}{1+v_2})]  \notag \\
&&\frac{1}{[u+\frac{1}{(1+v_1)(1+v_2)}]^3}(\ln\frac{M_c^2}{-p^2}%
-\gamma_\omega)  \notag \\
&=& -\frac{8ie^4\Gamma(4)}{(16\pi^2)^2} p^\mu p^\nu [\frac{5}{36}-\frac{1}{36%
}(\int^\infty_0\frac{dv_1}{1+v_1}+\int^\infty_0\frac{dv_2}{1+v_2})] (\ln%
\frac{M_c^2}{-p^2}-\gamma_\omega)  \notag \\
&=& -\frac{8ie^4\Gamma(4)}{(16\pi^2)^2} p^\mu p^\nu [\frac{5}{36}(\ln\frac{%
M_c^2}{-p^2}-\gamma_\omega)-\frac{1}{18}(\ln\frac{M_c^2}{-p^2}%
-\gamma_\omega) (\ln\frac{M_c^2}{-q_o^2}-\gamma_\omega)].
\end{eqnarray}
\indent For the more challenging part $\mathcal{M}^{(b)}_{301R}$, we follow
the approach already used in the derivation of $\mathcal{M}^{(b)}_{203R}$ in
order to obtain only the asymptotic results. We first focus on the region $%
v_1\rightarrow \infty$ while $v_2\rightarrow 0$, the expression for $%
\mathcal{M}^{(b)}_{301R}$ can be simplified to:
\begin{eqnarray}
\mathcal{M}^{(b)}_{301Rv_1} &\sim& -\frac{8ie^4\Gamma(4)}{(16\pi^2)^2}\cdot
\frac{1}{3}p^\mu p^\nu \int^1_0\prod^2_{i=1}dx_i \int^\infty_V \frac{dv_1}{%
(1+v_1)^3} \int^\infty_0 du \frac{x_2(1-x_2)}{(u+\frac{1}{1+v_1})^3} \ln%
\frac{x_2(1-x_2)}{u+\frac{1}{1+v_1}}  \notag \\
\end{eqnarray}
where the integration over $v_1$ is confined in the region from $V\gg 1$ to
infinity. It is easy to see that the simplified integral is essentially the
same as that for $\mathcal{M}^{(b)}_{203R}$, (\ref{b5_o-simp}), so we can
straightforwardly write down the result for $\mathcal{M}^{(b)}_{301R}$
\begin{eqnarray}  \label{b10_o}
\mathcal{M}^{(b)}_{301R} \sim \frac{8ie^4\Gamma(4)}{(16\pi^2)^2}\cdot\frac{1%
}{18}p^\mu p^\nu [\frac{13}{6}(\ln\frac{M_c^2}{-q^2_o(1+V)}-\gamma_\omega) -
\frac{1}{2}(\ln\frac{M_c^2}{-q^2_o}-\gamma_\omega)^2+\frac{1}{2}%
\alpha_\omega],
\end{eqnarray}
where we have already doubled the result for $\mathcal{M}^{(b)}_{301Rv_1}$
due to another asymptotic limit region $v_2\to \infty, v_1\to0$, which gives
the exactly same result.

A very similar calculation as $\mathcal{M}^{(b)}_{301}$ can give us the
results for $\mathcal{M}^{(b)}_{302}$,
\begin{eqnarray}
\mathcal{M}^{(b)}_{302} &=& -\frac{8ie^4\Gamma(4)}{(16\pi^2)^2}\cdot\frac{1}{%
3} g^{\mu\nu}p^2 \int^1_0\prod^2_{i=1}dx_i\int^\infty_0\prod^{2}_{i=1}\frac{%
dv_i}{(1+v_i)^3} \delta(1-\sum^2_{i=1}\frac{1}{1+v_i}) \int^\infty_0du
\notag \\
&& \frac{1}{[u+\frac{1}{(1+v_1)(1+v_2)}]^3} \{-3\frac{1} {[u+\frac{1}{%
(1+v_1)(1+v_2)}]^2}\frac{(x_2-x_1)^2}{(1+v_1)^4(1+v_2)^4}  \notag \\
&& +3\frac{(x_2-x_1)}{(1+v_1)^2(1+v_2)^2} (\frac{x_1}{1+v_1}+\frac{x_2}{1+v_2%
}-\frac{1}{2}) (\frac{1}{1+v_1}-\frac{1}{1+v_2}) \frac{1}{u+\frac{1}{%
(1+v_1)(1+v_2)}}  \notag \\
&& +\frac{1}{2}(\frac{x_1}{1+v_1}+\frac{x_2}{1+v_2})(1-\frac{x_1}{1+v_1}-%
\frac{x_2}{1+v_2}) [1-\frac{6}{(1+v_1)(1+v_2)}] \}(\ln\frac{M_c^2}{\mu_u^2}%
-\gamma_\omega)  \notag \\
&=& \mathcal{M}^{(b)}_{3020}+\mathcal{M}^{(b)}_{302R}.
\end{eqnarray}
\begin{eqnarray}  \label{b12_0}
\mathcal{M}^{(b)}_{3020} &=& -\frac{8ie^4\Gamma(4)}{(16\pi^2)^2}\cdot \frac{1%
}{3} g^{\mu\nu}p^2 \int^1_0\prod^2_{i=1}dx_i\int^\infty_0\prod^{2}_{i=1}%
\frac{dv_i}{(1+v_i)^3} \delta(1-\sum^2_{i=1}\frac{1}{1+v_i}) \int^\infty_0du
\notag \\
&& \frac{1}{[u+\frac{1}{(1+v_1)(1+v_2)}]^3} \{-3\frac{1} {[u+\frac{1}{%
(1+v_1)(1+v_2)}]^2}\frac{(x_2-x_1)^2}{(1+v_1)^4(1+v_2)^4}  \notag \\
&& +3\frac{(x_2-x_1)}{(1+v_1)^2(1+v_2)^2} (\frac{x_1}{1+v_1}+\frac{x_2}{1+v_2%
}-\frac{1}{2}) (\frac{1}{1+v_1}-\frac{1}{1+v_2}) \frac{1}{u+\frac{1}{%
(1+v_1)(1+v_2)}}  \notag \\
&& +\frac{1}{2}(\frac{x_1}{1+v_1}+\frac{x_2}{1+v_2})(1-\frac{x_1}{1+v_1}-%
\frac{x_2}{1+v_2}) [1-\frac{6}{(1+v_1)(1+v_2)}] \}(\ln\frac{M_c^2}{-p^2}%
-\gamma_\omega)  \notag \\
&=& -\frac{8ie^4\Gamma(4)}{(16\pi^2)^2}\cdot\frac{1}{3}g^{\mu\nu}p^2 (\ln%
\frac{M_c^2}{-p^2}-\gamma_\omega)[-\frac{43}{16\cdot9} +\frac{1}{24}%
(\int^\infty_0\frac{dv_1}{1+v_1}+\int^\infty_0\frac{dv_2}{1+v_2})]  \notag \\
&=& -\frac{8ie^4\Gamma(4)}{(16\pi^2)^2}\cdot\frac{1}{3}g^{\mu\nu}p^2 (\ln%
\frac{M_c^2}{-p^2}-\gamma_\omega)[-\frac{43}{16\cdot9} +\frac{1}{12}(\ln%
\frac{M_c^2}{-q_o^2}-\gamma_\omega)].
\end{eqnarray}
Note that the coefficient before $(\ln\frac{M_c^2}{-p^2}-\gamma_\omega)$ is
divergent, which means that $\mathcal{M}^{(b)}_{302}$ also contains
overlapping divergences. In order to deal with the resultant divergent
integral $\mathcal{M}^{(b)}_{302R}$, following the procedure used by $%
\mathcal{M}^{(b)}_{203R}$ and $\mathcal{M}^{(b)}_{301R}$, we need first to
simplify the integral in the region $v_1\rightarrow \infty,v_2\rightarrow 0$
\begin{eqnarray}
\mathcal{M}^{(b)}_{302Rv_1} &\sim& \frac{8ie^4\Gamma(4)}{(16\pi^2)^2}\cdot%
\frac{1}{6}g^{\mu\nu}p^2 \int^1_0\prod^2_{i=1}dx_i \int^\infty_V \frac{dv_1}{%
(1+v_1)^3} \int^\infty_0 du \frac{x_2(1-x_2)}{(u+\frac{1}{1+v_1})^3} \ln%
\frac{x_2(1-x_2)}{u+\frac{1}{1+v_1}}  \notag \\
&=&-\frac{8ie^4\Gamma(4)}{(16\pi^2)^2}\cdot\frac{1}{72}g^{\mu\nu}p^2 [\frac{%
13}{6}(\ln\frac{M_c^2}{-q^2_o(1+V)}-\gamma_\omega) - \frac{1}{2}(\ln\frac{%
M_c^2}{-q^2_o}-\gamma_\omega)^2+\frac{1}{2}\alpha_\omega].
\end{eqnarray}
In order to obtain $\mathcal{M}^{(b)}_{302R}$, the above result needs to be
doubled to take into account another contribution from the region $v_2\to
\infty, v_1\to 0$, which is given below:
\begin{equation}  \label{b12_o}
\mathcal{M}^{(b)}_{302R} \sim -\frac{8ie^4\Gamma(4)}{(16\pi^2)^2}\cdot \frac{%
1}{36}g^{\mu\nu}p^2[\frac{13}{6}(\ln\frac{M_c^2}{-q^2_o(1+V)}-\gamma_\omega)
- \frac{1}{2}(\ln\frac{M_c^2}{-q^2_o}-\gamma_\omega)^2+\frac{1}{2}%
\alpha_\omega].
\end{equation}

For the accomplishment of the computation of $\mathcal{M}^{(b)}_{30}$, we
have to calculate $\mathcal{M}^{(b)}_{303}$:
\begin{eqnarray}
\mathcal{M}^{(b)}_{303} &=& -\frac{8ie^4\Gamma(4)}{(16\pi^2)^2}\cdot \frac{2%
}{3}g^{\mu\nu}p^2 \int^1_0\prod^2_{i=1}dx_i\int^\infty_0\prod^{2}_{i=1}\frac{%
dv_i}{(1+v_i)^3} \delta(1-\sum^2_{i=1}\frac{1}{1+v_i}) \int^\infty_0du
\notag \\
&&\frac{1}{(1+v_1)^2(1+v_2)^2}(\ln\frac{M_c^2}{\mu_u^2}-\gamma_\omega)
\notag \\
&&\{\frac{1}{u+\frac{1}{(1+v_1)(1+v_2)}}[\frac{(x_2-x_1)^2}{(1+v_1)(1+v_2)} +%
\frac{x_1(1-x_1)}{1+v_1}+\frac{x_2(1-x_2)}{1+v_2}]  \notag \\
&&-\frac{1}{[u+\frac{1}{(1+v_1)(1+v_2)}]^2} \frac{(x_2-x_1)^2}{%
(1+v_1)^2(1+v_2)^2}\}  \notag \\
&=& \mathcal{M}^{(b)}_{3030}+\mathcal{M}^{(b)}_{303R},
\end{eqnarray}
where $\mathcal{M}^{(b)}_{3030}$ represents the part proportional to the
logarithmic divergence coming from $I^R_0$ while $\mathcal{M}^{(b)}_{303R}$
the rest parts. It is a direct exercise to obtain $\mathcal{M}^{(b)}_{3030}$
and the result is:
\begin{eqnarray}
\mathcal{M}^{(b)}_{3030} = -\frac{8ie^4\Gamma(4)}{(16\pi^2)^2}\cdot \frac{25%
}{18\cdot36}g^{\mu\nu}p^2(\ln\frac{M_c^2}{-p^2}-\gamma_\omega),
\end{eqnarray}
while part $\mathcal{M}^{(b)}_{303R}$ can be proven finite. So the
divergence structure of $\mathcal{M}^{(b)}_{303}$ is given by:
\begin{eqnarray}  \label{b8}
\mathcal{M}^{(b)}_{303} \sim -\frac{8ie^4\Gamma(4)}{(16\pi^2)^2}\cdot \frac{%
25}{18\cdot36}g^{\mu\nu}p^2(\ln\frac{M_c^2}{-p^2}-\gamma_\omega),
\end{eqnarray}

For the part $\mathcal{M}^{(b)}_{3R}$, we can prove that most terms are
finite for the integration of UVDP or Feynman parameters, except for the
following one:
\begin{eqnarray}  \label{b11}
\mathcal{M}^{(b)}_{3R1} &=& -\frac{8ie^4\Gamma(4)}{16\pi^2}\cdot\frac{1}{3}%
(g^{\mu\nu}p^2-2p^\mu
p^\nu)\int^1_0\prod^2_{i=1}dx_i\int^\infty_0\prod^{2}_{i=1}\frac{dv_i}{%
(1+v_i)^3} \delta(1-\sum^2_{i=1}\frac{1}{1+v_i})  \notag \\
&&\int^\infty_0du [\frac{1}{(1+v_1)(1+v_2)} -(\frac{x_1}{1+v_1}+\frac{x_2}{%
1+v_2})(1-\frac{x_1}{1+v_1}-\frac{x_2}{1+v_2})]  \notag \\
&& \frac{1}{[u+\frac{1}{(1+v_1)(1+v_2)}]^3} \frac{1}{16\pi^2}\frac{1}{%
2\mu_u^2}\mu_u^2  \notag \\
&=& -\frac{8ie^4\Gamma(4)}{(16\pi^2)^2}\cdot\frac{1}{6}(g^{\mu\nu}p^2-2p^\mu
p^\nu)\int^\infty_0\prod^2_{i=1}\frac{dv_i}{(1+v_i)^2} \delta(1-\sum^2_{i=1}%
\frac{1}{1+v_i})  \notag \\
&& [\frac{5}{12}-\frac{1}{12}(1+v_1)(1+v_2)]  \notag \\
&=& -\frac{8ie^4\Gamma(4)}{(16\pi^2)^2}\cdot\frac{1}{6}(g^{\mu\nu}p^2-2p^\mu
p^\nu) [\frac{5}{12}-\frac{1}{6}(\ln\frac{M_c^2}{-q_o^2}-\gamma_\omega)]
\notag \\
&\sim& -\frac{8ie^4\Gamma(4)}{(16\pi^2)^2}\cdot(g^{\mu\nu}p^2-2p^\mu p^\nu)(-%
\frac{1}{36})(\ln\frac{M_c^2}{-q_o^2}-\gamma_\omega)
\end{eqnarray}

By putting the divergence parts of the terms Eqs.(\ref{b2}), (\ref{b1}), (%
\ref{b3}), (\ref{b4}), (\ref{b6}), (\ref{b5_0}), (\ref{b5_o}), (\ref{b7}), (%
\ref{b10_0}), (\ref{b10_o}), (\ref{b12_0}), (\ref{b12_o}), (\ref{b8}), (\ref%
{b11}) together, we finally arrive at the divergence behavior of the diagram
$\mathcal{M}^{(b)}$.
\begin{eqnarray}  \label{b_result}
\mathcal{M}^{(b)} &\sim& -\frac{8ie^4}{(16\pi^2)^2}\{g^{\mu\nu}M_c^2(\ln%
\frac{M_c^2}{-q_o^2}-\gamma_\omega) +\frac{1}{6}g^{\mu\nu}p^2(\ln\frac{M_c^2%
}{-q_o^2}-\gamma_\omega)  \notag \\
&&+(g^{\mu\nu}p^2-p^\mu p^\nu)\cdot[\frac{1}{3}(\ln\frac{M_c^2}{-p^2}%
-\gamma_\omega)(\ln\frac{M_c^2}{-q_o^2}-\gamma_\omega)+ \frac{19}{18}(\ln%
\frac{M_c^2}{-p^2}-\gamma_\omega)]  \notag \\
&&-\frac{1}{3}(\ln\frac{M_c^2}{-q_o^2}-\gamma_\omega) +\frac{1}{3}[\frac{13}{%
6}(\ln\frac{M_c^2}{-q_o^2(1+V)}-\gamma_\omega)- \frac{1}{2}(\ln\frac{M_c^2}{%
-q^2_o}-\gamma_\omega)^2+\frac{1}{2}\alpha_\omega] \}  \notag \\
\end{eqnarray}


\section{Derivation of the Integration $\protect\int^\infty_0 \frac{d v_2}{%
1+v_2}\ln(1+v_2)$}

Recall that in the above derivation of the two-loop massless QED photon
vacuum polarization diagrams we encountered a new parameter integral
\begin{eqnarray}  \label{newI}
\int^\infty_0 \frac{d v_2}{1+v_2}\ln(1+v_2)
\end{eqnarray}
which does not appear at one-loop level. Thus, for completion of our
calculation, in this section we shall derive its regulated result with the
LORE method.

With the same philosophy as before, we would like to transform this UVDP
parameter integral to a momentum-like one. In order to do so, we just need
to multiply a free energy scale $-q_o^2$ in the numerator and denominator
simultaneously,
\begin{eqnarray}
\int^\infty_0 \frac{d v_2}{1+v_2}\ln(1+v_2) &=& \int^\infty_0 \frac{d q^2}{%
q^2-q_o^2}\ln(q^2-q_o^2)- \int^\infty_0 \frac{d q^2}{q^2-q_o^2}\ln (-q_o^2)
\notag \\
&=& I_1-I_2 ,
\end{eqnarray}
where we have defined $q^2=-q_o^2 v_2$ and separated the integral into two
parts $I_1$ and $I_2$. The integration $I_2$ can be easily worked out with
the LORE method since it is exactly the one encountered in the one-loop
calculations. The result of $I_2$ is
\begin{eqnarray}
I_2 = \ln (-q_o^2) [\ln\frac{M_c^2}{\mu_q^2}-\gamma_\omega+y_0\big(\frac{%
\mu_q^2}{M_c^2}\big)]
\end{eqnarray}
where we define $\mu_q^2\equiv\mu_s^2-q_o^2$.

The integration $I_1$ is really new and requires us to calculate more
carefully. In the following, we shall give the detailed derivation of this
integration in the LORE method. As usually done in the LORE method, we need
to apply a series of regulators to the integral:
\begin{eqnarray}
I_1 &=& \lim_{N\to\infty}\sum^N_{l=0} c_l^N \frac{d q^2}{q^2+\hat{\mathcal{M}%
}_l^2}\ln(q^2+\hat{\mathcal{M}}^2_l),
\end{eqnarray}
where we defined $\hat{\mathcal{M}}^2_l=-q_o^2+\mu_s^2+l M_R^2=\mu_q^2+ l
M_R^2$. With the help of the following equality:
\begin{eqnarray}
\lim_{N\to\infty}\sum^N_{l=0}c_l^N l^n=0\quad \mbox{with} \quad n=
0,1,...,N-1,
\end{eqnarray}
we can easily work out the above integration:
\begin{eqnarray}
I_1 &=& \lim_{N\to\infty}\sum^N_{l=0} c_l^N (-\frac{1}{2}) \ln^2\hat{%
\mathcal{M}}^2_l = (-\frac{1}{2})\lim_{N\to\infty}\sum^N_{l=0} c_l^N
\ln^2(\mu^2_q+l M^2_R)  \notag \\
&=& (-\frac{1}{2})\{\ln^2\mu^2_q
+\lim_{N\to\infty}\sum^N_{l=1}c_l^N\ln^2(\mu^2_q+ l M_R^2)\}  \notag \\
&=& (-\frac{1}{2})\{\ln^2\mu^2_q +\lim_{N\to\infty}\sum^N_{l=1}c_l^N[\ln(1+
\frac{\mu_q^2}{l M_R^2})+ \ln l +\ln M_R^2]^2\}  \notag \\
&=& (-\frac{1}{2})\{\ln^2\mu^2_q +\lim_{N\to\infty}\sum^N_{l=1}c_l^N [\ln
M_c(N)^2 -\gamma_\omega(N)+\ln l +\sum^N_{l^\prime=1} c_{l^\prime}^N \ln
l^\prime + \ln(1+ \frac{\mu_q^2}{l M_R^2})]^2\}  \notag \\
&=& \frac{1}{2}\{(\ln M_c^2-\gamma_\omega)^2 -\ln^2\mu^2_q -\alpha_\omega
+2\ln \mu^2_q\cdot y_0(\frac{\mu_q^2}{M_c^2}) + y_0^\prime(\frac{\mu_q^2}{%
M_c^2}) \}
\end{eqnarray}
where we have introduced two series:
\begin{eqnarray}
M_c^2 (N) \equiv \sum^N_{l=1}c^N_l (l\ln l)M_R^2, \quad \mbox{and} \quad
\gamma_\omega(N) \equiv \sum^N_{l=1} c^N_l \ln l + \ln[\sum^N_{l^%
\prime}c^N_{l^\prime} (l^\prime \ln l^\prime)],
\end{eqnarray}
both of which in the limit of $N\to \infty$ can be shown \cite{wu1} to have
finite limits
\begin{eqnarray}
M_c^2 = \lim_{N\to \infty} M_c^2(N),\quad \mbox{and} \quad
\gamma_\omega=\lim_{N\to \infty} \gamma_\omega(N).
\end{eqnarray}
We also define a new function
\begin{equation}
y_0^\prime (\frac{\mu_q^2}{M_c^2}) = -\lim_{N,M_R^2} \sum_{l=1}^{N} c_l^N
\ln(1+\frac{\mu_q^2}{l M_R^2})[2(\ln l +\ln \frac{M_c^2(N)}{\mu_q^2}
-\gamma_\omega(N) +\sum^N_{l^\prime=1} c^N_{l^\prime}\ln l^\prime) + \ln(1+%
\frac{\mu_q^2}{l M_R^2})],
\end{equation}
which can be seen that when the UV scale $M_c^2$ tends to infinity the
function $y_0^\prime (\frac{\mu_q^2}{M_c^2})$ vanishes since the expansion
of $\ln(1+\frac{\mu_q^2}{l M_R^2})$ is of the order of $O(\frac{\mu_q^2}{%
M_R^2})$. The constant $\alpha_\omega$ is defined as
\begin{equation}
\alpha_\omega \equiv \lim_{N\to\infty}[\sum^N_{l=1} c_l^N (\ln
l)^2+(\sum^N_{l=1}c_l^N \ln l)^2].
\end{equation}
It can be shown from our numerical calculation that $\alpha_\omega$ is
finite and is $\alpha_\omega=1.62931 ...$.

By combining the expression $I_1$ and $I_2$, we can give our final
regularized result
\begin{eqnarray}
&& \int^\infty_0 \frac{d v_2}{1+v_2}\ln(1+v_2) = I_1-I_2  \notag \\
&=& \frac{1}{2}[(\ln\frac{M_c^2}{\mu_q^2}-\gamma_\omega)^2 +2\ln\frac{\mu_q^2%
}{-q_o^2}(\ln\frac{M_c^2}{\mu_q^2}-\gamma_\omega)-\alpha_\omega + 2\ln\frac{%
\mu_q^2}{-q_o^2}y_0(\frac{\mu_q^2}{M_c^2})+y_0^\prime(\frac{\mu_q^2}{M_c^2}%
)].
\end{eqnarray}
When we set the IR scale $\mu_s^2$ to 0 which can be viewed as the IR cutoff
in the LORE method, then the regulated integral can be simplifies to:
\begin{eqnarray}
\int^\infty_0 \frac{d v_2}{1+v_2}\ln(1+v_2) = \frac{1}{2}[(\ln\frac{M_c^2}{%
-q_o^2}-\gamma_\omega)^2 -\alpha_\omega +y_0^\prime(\frac{-q_o^2}{M_c^2})].
\end{eqnarray}
If we further take the limit $M_c\to \infty$, the last term $y_0^\prime(%
\frac{-q_o^2}{M_c^2})\to0$ also, which gives the result we use in our
previous calculations.



\vspace*{0.5cm}

\begin{thebibliography}{99}

\bibitem{Huang} D.~Huang and Y.~-L.~Wu,
Eur.\ Phys.\ J.\ C \textbf{72}, 2066 (2012) [arXiv:1108.3603 [hep-ph]].


\bibitem{'tHooft:1972fi} G.~'t Hooft and M.~J.~G.~Veltman,
Nucl.\ Phys.\ B \textbf{44}, 189 (1972). 

\bibitem{wu1} Y.~L.~Wu,
Int.\ J.\ Mod.\ Phys.\ A \textbf{18}, 5363 (2003) [arXiv:hep-th/0209021].

\bibitem{wu2} Y.~L.~Wu,
Mod.\ Phys.\ Lett.\ A \textbf{19}, 2191 (2004) [arXiv:hep-th/0311082].

\bibitem{bjor} J. Bjorken and S. Drell, \textit{USA: McGraw-Hill Book
Company (1965) 396 p}


\bibitem{Peskin:1995ev} M.~E.~Peskin and D.~V.~Schroeder,
\textit{Reading, USA: Addison-Wesley (1995) 842 p}

\bibitem{Itzykson:1980rh} C.~Itzykson and J.~B.~Zuber,
\textit{New York, USA: Mcgraw-hill (1980) 705 P.(International Series In
Pure and Applied Physics)}

\bibitem{Jost: 1950} R.~Jost and J. ~M.~Luttinger, \textit{Helv. Phys. Acta,
23, 201 (1950)}
\end{thebibliography}
\end{document}